\documentclass[preprint]{aastex}
\usepackage{epsf}

\def\linebreak{\hfil\break}

\def\
{\hfil\linebreak}

\def\degree{\ifmmode {^\circ}\else {$^\circ$}\fi}
\def\mum{\ifmmode {\rm \mu {\rm m}}\else $\rm \mu {\rm m}$\fi}
\def\arcsec{\ifmmode ^{\prime \prime}\else $^{\prime \prime}$\fi}

\def\inch{\ifmmode ^{\prime \prime}\else $^{\prime \prime}$\fi}
\def\arcmin{\ifmmode ^{\prime}\else $^{\prime}$\fi}

\def\msun{\ifmmode {\rm M_{\odot}}\else $\rm M_{\odot}$\fi}
\newbox\grsign \setbox\grsign=\hbox{$>$} \newdimen\grdimen \grdimen=\ht\grsign
\newbox\simlessbox \newbox\simgreatbox
\setbox\simgreatbox=\hbox{\raise.5ex\hbox{$>$}\llap
     {\lower.5ex\hbox{$\sim$}}}\ht1=\grdimen\dp1=0pt
\setbox\simlessbox=\hbox{\raise.5ex\hbox{$<$}\llap
     {\lower.5ex\hbox{$\sim$}}}\ht2=\grdimen\dp2=0pt

\def\simless{\mathrel{\copy\simlessbox}}

\begin{document}

\title{Collisional Cascades in Planetesimal Disks 
I. Stellar Flybys}
\vskip 7ex
\author{Scott J. Kenyon}
\affil{Smithsonian Astrophysical Observatory,
60 Garden Street, Cambridge, MA 02138} 
\email{e-mail: skenyon@cfa.harvard.edu}

\author{Benjamin C. Bromley}
\affil{Department of Physics, University of Utah, 
201 JFB, Salt Lake City, UT 84112} 
\email{e-mail:bromley@physics.utah.edu}
%
\received{1 August 2001}
\accepted{19 Novmber 2001}

%
%

\begin{abstract}

We use a new multiannulus planetesimal accretion code to investigate 
the evolution of a planetesimal disk following a moderately close 
encounter with a passing star.  The calculations include fragmentation, 
gas and Poynting-Robertson drag, and velocity evolution from dynamical 
friction and viscous stirring.  We assume that the stellar encounter 
increases planetesimal velocities to the shattering velocity, initiating 
a collisional cascade in the disk.  
During the early stages of our calculations, erosive collisions damp 
particle velocities and produce substantial amounts of dust.  For a wide 
range of initial conditions and input parameters, the time evolution 
of the dust luminosity follows a simple relation,
$L_d/L_{\star} = L_0 / [\alpha + (t/t_d)^{\beta}]$.  The maximum 
dust luminosity $L_0$ and the damping time $t_d$ depend on the 
disk mass, with $L_0 \propto M_d$ and $t_d \propto M_d^{-1}$.  
For disks with dust masses of 1\% to 100\% of the `minimum mass solar 
nebula' (1--100 $M_{\oplus}$ at 30--150 AU), our calculations yield 
$t_d \sim$ 1--10 Myr, $\alpha \approx$ 1--2, $\beta$ = 1, and dust 
luminosities similar to the range observed in known `debris disk' 
systems, $L_0 \sim 10^{-3}$ to $10^{-5}$.  Less massive disks produce 
smaller dust luminosities and damp on longer timescales.  Because 
encounters with field stars are rare, these results imply that 
moderately close stellar flybys cannot explain collisional cascades 
in debris disk systems with stellar ages of $\sim$ 100 Myr or longer.

\end{abstract}

\subjectheadings{planetary systems -- solar system: formation -- 
stars: formation -- circumstellar matter}

\section{INTRODUCTION}

Many nearby main sequence stars have thermal emission from cold dust 
\citep{bac93,art97,lag00}.  
These `debris disk' systems often have a disk-like or ring morphology 
at optical or infrared wavelengths, with typical radii of 50--1000 AU 
\citep[e.g.,][]{smi84,gre98,hol98,jay98,koe98,sch99,wei99}.  
The near-IR and far-IR excess emission is a small fraction of the stellar 
luminosity; the ratio of the total dust luminosity to the stellar 
luminosity is $L_d/L_{\star} \sim 10^{-5}$ 
to $10^{-2}$ in most systems \citep{aum84,faj99,hab99,spa01}. 
Recent studies indicate that the dust luminosity declines on timescales
of 10--100 Myr \citep{bar99,hab99,so00b}.  
\citet{kal98} and \citet{spa01} propose power-law relations for this 
decline, $L_d \propto t^{-1}$ \citep{kal98} and
$L_d \propto t^{-2}$ \citep{spa01}.

In current theories, dynamic processes produce dust emission in a 
debris disk.  If the disk contains small dust grains with sizes of 
1--100 $\mu$m and a total mass in small grains of $\sim$ 0.01 
$M_{\oplus}$\footnote{1 $M_{\oplus} = 6 \times 10^{27}$ g is the mass 
of the earth}, radiative transfer models can explain both the scattered-light 
images and the spectral energy distributions of most systems 
\citep[e.g.,][Greaves, Mannings, \& Holland 2000b]{bac93,art97}.  However,
radiation pressure and Poynting-Robertson drag remove small grains from 
the disk on short timescales, $\sim$ 1--10 Myr, compared to the age of 
a typical system.  Collisions between larger bodies can replenish the 
small grain population when their relative velocities are large enough
to begin a `collisional cascade,' where 1--10 km planetesimals are
ground down into smaller and smaller bodies. If planetesimal velocities 
are close to the `shattering velocity' of $\sim$ 200 m s$^{-1}$ and 
if the mass in 1--10 km planetesimals is $\sim$ 100 $M_{\oplus}$, 
collisional cascades can maintain the observed small grain population over 
timescales of $\sim$ 100 Myr or longer \citep{hab99,gr00b,kb01}.  

At least two mechanisms can produce large collision velocities in 
a debris disk.  During the early evolution of the disk, there is a 
strong coupling between solid bodies and the gas; circularization is 
efficient and collision velocities are low.  As solid bodies in the disk 
merge and grow, impulsive encounters with passing stars or stirring
by large planets embedded in the disk can increase collision velocities
\citep{art89,mou97,ida00,kal00}.  Kenyon \& Bromley (2001) show that 
embedded planets with radii of 500 km or larger can dynamically heat 
smaller bodies to the shattering velocity on timescales of 10--100 Myr.  
Collisions between 1--10~km objects can then replenish the small grain
population for $\sim$ 500 Myr or longer \citep[see also][]{ken02}.  
Ida et al. (2000) and 
Kalas et al. (2000) use $n$-body simulations to demonstrate that 
close encounters between the disk and a passing star can excite 
large particle velocities in the disk.  Because the growth of large 
planets in the outer disk is unlikely, these `stellar flybys' are 
an attractive way to induce collisional cascades in the large disks 
observed in $\beta$ Pic and other nearby stars.  The evolution of a 
particle disk after the stellar flyby is uncertain.  Collisions of 
small bodies in the disk may damp planetesimal velocities and thus halt 
the collisional cascade.  \citet{kl99} derive damping times of 1--10 Myr 
for 1--100 m planetesimals in the Kuiper Belt.  If this timescale is
typical, then stellar flybys can explain debris disks only if the
encounters are frequent.

Here we investigate the evolution of a planetesimal disk following 
a moderately close
stellar flyby.  We use a new multi-annulus planetesimal evolution 
code to compute the collisional and dynamical evolution of small 
bodies in an extended disk surrounding an intermediate mass star.  
Calculations for disks with dust masses comparable to the `minimum 
mass solar nebula' ($\sim$ 100 $M_{\oplus}$ of solid material at 
30--150 AU) show that two body collisions can damp planetesimal 
velocities on timescales of 1--10 Myr.  Less massive disks damp on 
longer timescales.  Small dust grains generated in the collisional 
cascade produce a significant luminosity, with
$L_d/L_{\star} \sim 5 \times 10^{-2}$ for a minimum mass solar nebula. Less
massive disks yield smaller dust luminosities.  These results suggest 
that stellar flybys cannot initiate luminous, long-lived collisional
cascades in a planetesimal disk.  Unless all known debris disk systems 
have had a stellar encounter in the past 10--100 Myr, a nearby binary 
companion star or planets embedded in the disk probably initiated the 
collisional cascades in these systems.

We outline the model in \S2, describe the calculations in \S3,
and conclude with a brief discussion in \S4.

\section{THE MODEL}

We treat planetesimals as a statistical ensemble of masses with a
distribution of horizontal and vertical velocities about a Keplerian
orbit \citep[][1999, and references therein]{saf69,wet89,kl98}.  
The model grid contains $N$ concentric annuli centered at heliocentric
distances $a_i$.  Each annulus has an inner radius at
$a_i - \delta a_i/2$ and an outer radius at $a_i + \delta a_i/2$.
The midpoint of the model grid is at a heliocentric distance $a_{mid}$.
Calculations begin with a differential mass distribution $n(m_{ik}$)
of bodies with horizontal and vertical velocities $h_{ik}(t)$ and $v_{ik}(t)$.  
 
To evolve the mass and velocity distributions in time, we solve the 
coagulation and the Fokker-Planck equations for an ensemble of masses 
undergoing inelastic collisions, drag forces, and dynamical friction 
and viscous stirring.  We approximate collision cross-sections using
the particle-in-a-box formalism and treat gas and Poynting-Robertson 
drag with simple analytic formulae (see the Appendix).  
\citet{kb01} outline our treatment of elastic gravitational interactions
\citep[see also][]{spa91,wei97}.

We allow four collision outcomes
\citep{gre78,dav85,bar93,wet93}:

\vskip 2ex
\noindent
1. Mergers -- two bodies collide and merge into a single object
with no debris;

\noindent
2. Cratering -- two bodies merge into a single object but produce
debris with mass of 20\% or less of the mass of the merged object;

\noindent
3. Rebounds -- two bodies collide and produce some debris but do
not merge into a single object; and

\noindent
4. Disruption -- two bodies collide and produce debris with a mass
comparable to the mass of the two initial bodies.
\vskip 2ex
\noindent
The appendix of \citet{kl99} describes the algorithm for specifying
the outcome as a function of the collision velocity, the tensile
strength of the colliding bodies $S_0$, and other physical parameters.
For the calculations described in this paper, all collisions result
in debris; rebounds are unimportant.  For most calculations, we 
adopt $S_0 = 2 \times 10^6$ erg g$^{-1}$ and a crushing energy 
$Q_c = 5 \times 10^7$ erg g$^{-1}$, appropriate for icy objects at large 
distances from the central star \citep[see][and references therein]{kl99}.

The initial conditions for these calculations are appropriate for a
disk with an age of $\sim$ 10 Myr.  We consider systems of $N$ annuli in 
disks with $a_{mid}$ = 35--140 AU and $\delta a_i/a_i$ = 0.01--0.025.  
The disk is composed of small planetesimals with radii of $\sim$ 
1--100 m (see below).  The particles have an initial mass distribution 
$n_i(m_k)$ in each annulus and begin with eccentricity $e_0$ and inclination 
$i_0 = e_0/2$.  Most of our models have $e_0$ independent of $a_i$; some 
models have a constant initial horizontal velocity in each annulus.
We consider moderately strong perturbations with $e_0$ = 0.01--0.06.
Very close stellar flybys can produce $e_0 \gtrsim$ 0.1
\citep[e.g.,][]{ida00,kal00,kob01}; our adopted $e_0$ is 10--100 
times larger than for a cold planetesimal disk, where 
$e_0 \lesssim 10^{-3}$ \citep{kl99}.  We assume a power law variation 
of the initial surface density of solid material with heliocentric 
distance, $\Sigma_i = \Sigma_0 (a_i/a_0)^{-n}$; models with $n$ = 1.5 
are `standard'.  

We assume a stellar flyby instantaneously raises the orbital eccentricity 
and inclination of all planetesimals in the disk.
Although not strictly accurate, this approximation is valid if the
encounter between the passing star and the disk is short compared to
the damping time.  Any passing star that is not bound to the central 
star of the disk satisfies this requirement \citep{lar97,mou97}.  
A bound companion star with an orbital semi-major axis of $\sim$ 1000 AU 
or less can excite large planetesimal velocities over long timescales 
and may counteract damping.  Embedded planets with radii of 500 km or 
larger can also counteract damping \citep{kb01}.
We plan to consider both situations in future papers. 

Our calculations follow the time evolution of the mass and velocity 
distributions of objects with a range of radii, $r_{ik} = r_{min}$
to $r_{ik} = r_{max}$.  The upper limit $r_{max}$ is always larger
than the largest object in each annulus.  To save computer time, we 
do not consider small objects which do not affect significantly the
dynamics and growth of larger objects.  Usually this limit is
$r_{min}$ = 10 cm.  Erosive collisions produce objects with $r_{ik}$
$< r_{min}$.  These particles are `lost' to the model grid.  For
most conditions, lost objects are more likely to be ground down into 
smaller objects than to collide with larger objects in the grid.
Thus, our neglect of lost particles with $r_{ik} < r_{min}$ is a
reasonable approximation.

We do not consider planetesimals in the inner disk. Test calculations
show that collisions damp particle velocities on short timescales,
$t_d \lesssim 10^3$ yr at $a_i \lesssim$ 5 AU.  For a stellar flyby with
an encounter velocity of 1--10 km s$^{-1}$, these timescales are
comparable to the duration of the flyby \citep[see][]{lar97,mou97}.  
Our approximation of an instantaneous impulse to the disk in then
invalid.  The short damping timescales for the inner disk also
suggest that the chances of finding a disk in this state are small.  We 
therefore consider the evolution of the outer disk in these calculations.

For a cold disk of planetesimals with $e_0 \lesssim 10^{-3}$, the 
initial conditions of our calculations are sufficient to produce 
1000 km or larger planets at 30 AU on timescales of $\sim$ 10 Myr 
for a minimum mass solar nebula \citep[e.g.,][]{st97a,kl99,ken99}.  
When a stellar flyby raises particle eccentricities by factors of 10--100,
collisions between planetesimals promote erosion instead of growth by 
mergers.  Our goal is to identify perturbations where collisional
damping can reduce particle eccentricities and promote growth before 
erosive collisions reduce planetesimals to dust. 

\section{CALCULATIONS}

\subsection{Cascades in Rings}

To understand the response of a planetesimal disk to a stellar flyby,
we begin by considering a small portion of a disk surrounding a 
3 \msun~star.  The ring consists of bodies with $r_{min}$ = 10 cm to
$r_{max}$ in 16 annuli with $\delta a_i/a_i$ = 0.01.  Each body has
$e_0$ = 0.02 and a mass density of 1.5 g cm$^{-3}$.  
The mass spacing factor is $\delta_1 = m_{k+1}/m_k$ = 2 for all batches;
the initial mass distribution has equal mass per batch. 
The surface density of solid material is $\Sigma_d(a)$ = 0.15
$x (a / {\rm 35 ~ AU})^{-3/2}$ g~cm$^{-2}$, where $x$ is a 
dimensionless constant.  Models with $x$ = 1--2 have a total mass
in solid material similar to the `minimum mass solar nebula' 
at 2--20 AU \citep{wei77a,hay81}.  

We calculate the collisional evolution for a 
set of models with $x$ = $10^{-5}$ to $1$ and $r_{max}$ = 
10~m to 1000~km at $a_{mid}$ = 35 AU, 70 AU, and 140 AU.  
For computational speed, we adopt the \citet{wet93} fragmentation 
algorithm. Calculations with the \citet[][1994]{dav85} algorithm 
evolve on a faster timescale when the collisional debris receives
more kinetic energy per unit mass than the merged object 
\citep[see][]{kl99}.  Collisional debris rarely has less kinetic
energy per unit mass than the merged object \citep[][1994]{dav85}.
The fragmentation parameters for our standard models are
$Q_c$ = $5 \times 10^7$ erg g$^{-1}$ (crushing energy),
$S_0$ = $2 \times 10^6$ erg g$^{-1}$ (impact strength), and
$V_f$ = 1 m s$^{-1}$ (the minimum velocity for cratering).

Figures 1--2 outline the evolution of a model with $x$ = 0.1 
and $r_{max}$ = 10 m at $a_{mid}$ = 35 AU.  The 16 annuli in this 
standard model contain an initial mass of $ 4 \times 10^{27}$ g 
(0.67 $M_{\oplus}$).  Erosive collisions
dominate the early stages of this calculation.  Collisions
between small bodies with $r_k$ = 10--100 cm are completely
disruptive; collisions between the largest bodies yield some 
growth and substantial cratering.  Throughout the first $10^4$ yr 
of the calculation, the initial `mass loss rate' is large: $\sim$
$ 10^{22}$ g yr$^{-1}$ is lost to particles with 
$r_k < r_{min}$ = 10 cm.
This rate falls uniformly with time thereafter. Disruptive 
collisions between the `lost' particles probably produce 
micron-sized grains which are ejected from the system on 
short timescales, $\lesssim 10^5$ yr \citep[e.g.,][]{kri00}.  

As the calculation continues past $10^4$ yr, collisional damping
reduces the velocities of the mid-sized particles with $r_k \sim$ 1 m.  
Collisions between these and the largest bodies in the grid produce 
growth and some debris.  The debris produced from these and other 
collisions flattens the cumulative number distribution from the 
initial $N_C \propto m_i^{-1}$ to $N_C \propto m_i^{-0.83}$.  For 
the rest of the calculation, the power law slope of the number 
distribution for the small bodies remains close to $-0.83$ in each 
annulus.  This slope is a standard result of coagulation when 
collisions produce debris \citep{doh69,wet93,wil94,tan96,kl99}.  As 
collisional damping continues to reduce the velocities of the mid-sized 
bodies, the largest bodies grow more rapidly.  By $t$ = 1 Myr, 
the largest bodies have $r_k$ = 25 m (Figure 1, middle left panel).  
Although the largest and smallest bodies retain a large fraction 
of their initial eccentricities, the orbits of the mid-sized bodies 
are a factor of $\sim$ 4 less eccentric than their initial orbits 
(Figure 1; middle right panel).  As growth proceeds, dynamical friction 
considerably reduces the velocities of the largest objects.  Collisions 
damp the velocities of the smaller bodies.  By $t$ = 3 Myr, the 
largest objects have $r_k$ = 100 m and most objects have 
$e_k \le 0.1 ~ e_0$ (Figure 1, lower panels).  From previous results,
the largest bodies soon begin `runaway growth' and reach sizes 
of 10--100 km or larger on timescales of $\sim$ 30--100 Myr 
\citep{gre78,kl99,ken99}.  Because the production of debris -- already 
below $\sim 10^{21}$ g yr$^{-1}$ -- drops rapidly during runaway
growth, we stopped this calculation at $t$ = 10 Myr.

Figure 2 shows the time evolution of two `observables', the radial 
optical depth and the fraction of stellar luminosity intercepted
and reprocessed by solid material in the grid, $L/L_{\star}$.  We 
follow \citet{ken99} and calculate the radial optical depth $\tau_d$ 
of particles in the grid using the geometric optics limit.  
For particles with $r_k <$ 10 cm, we estimate a minimum optical 
depth $\tau_{min}$ assuming Poynting-Robertson drag is efficient 
at removing the smallest objects and a maximum optical depth 
$\tau_{max}$ assuming Poynting-Robertson drag removes none of 
the smallest objects.  The dust luminosity follows directly from the
solid angle of the dusty annulus as seen from the central star,
$L/L_{\star}$ = $\tau H_d / a$, where $H_d$ is the scale height of the
dust.  The appendix describes our derivation of the radial optical depth 
and the reprocessed stellar luminosity in more detail 
\citep[see also][]{kri00}.

At $t$ = 0, the optical depth is large because the dust production 
rate is large.  The optical depth of the smallest grains,
$\tau_{min}$ and $\tau_{max}$, falls with 
the declining dust production rate and drops by $\sim$ two orders of 
magnitude by $t$ = 10 Myr.  For $t <$ 1 Myr, the optical depth of 
the mid-sized bodies grows with time.  Collisional damping reduces 
the scale height of these particles, which increases their number density 
and thus their radial optical depth.  By $t = $ 10 Myr, the largest 
bodies begin to grow rapidly and $\tau_d$ falls back to its initial 
level. The optical depth continues to drop as runaway growth 
concentrates more and more mass in the largest bodies.

The evolution of the dust luminosity depends on the evolution of
the optical depth and the dust scale height\footnote{We do not attempt 
to model the evolution of the dust luminosity during the stellar flyby.
Because we assume a size distribution for small particles, 
the luminosity at $t$ = 0 in our models assumes that these particles
reach the collisional size distribution, $N_C \propto m_i^{-0.83}$,
instantaneously.  In a more realsitic calculation, the timescale
to achieve this size distribution, $\sim$ $10^4$ yr to $10^5$ yr,
is short compared to the damping time.  Thus, our model provides a 
reasonably good approximation to the luminosity evolution following
the flyby, but it makes no prediction for the change in $L/L_0$ 
resulting from the flyby.}
Because collisional damping reduces particle velocities throughout the 
calculation, the dust scale height falls monotonically with time.  Unless 
the optical depth rises substantially, decreasing dust scale heights 
result in a declining dust luminosity.  The small rise in $\tau_d$ for 
$t < $ 1 Myr does not compensate for the decline in the scale 
height: the dust luminosity thus declines with time.  At $t = $ 0, 
the dust luminosity roughly equals the observed luminosities of 
debris disk systems with ring-like geometries, $L/L_{\star} \sim$ 
$10^{-3}$ \citep[e.g.,][]{au99b,gre98,jay98,koe98}.  
By the end of the calculation, our most optimistic
estimate of the dust luminosity falls below the smallest observed
luminosity in known debris disk systems.

Although our predicted luminosities are close to those observed,
the radial optical depths in our calculations, $\sim 10^{-1}$ for the
small particles and $\sim 10^{-2}$ large particles in the grid,
are significantly larger than values of $\sim 10^{-3}$ or less
typically quoted for debris disk systems.  Published optical 
depths are usually derived from the ratios of the dust luminosity
to the stellar luminosity and assume the dust is arranged in
a spherical shell, $L/L_{\star}$ = $\tau (1 - \omega)$, where 
$\omega$ is the albedo.  We adopt a more appropriate ring geometry; 
the dust luminosity is $L/L_{\star}$ = $\tau (1 - \omega) H / a$.
For rings with $H / a \sim$ 0.01--0.1, the radial optical depths 
must be $\sim$ 10--100 times larger through the midplane of a ring
than through a spherical shell.  With this scaling, our optical
depth estimates agree with published estimates for debris disk
systems.

The luminosity evolution for each curve in Figure 2 is well-fit by 
a simple function
\begin{equation}
L/L_{\star} = \frac{L_0}{\alpha + (t/t_d)^{\beta}} ~ ,
\end{equation}
where $L_0$ is the initial dust luminosity.  Because collisional 
damping reduces the eccentricities of the mid-sized particles by 
a factor of $\sim$ 2 at $t = t_d$, we call $t_d$ the `damping time.'  
At late times, the luminosity follows a power law with index $\beta$.
We derive $\beta = 1.35 \pm 0.02$ for the luminosity of small particles,
$L_{min}$ and $L_{max}$,
and $\beta = 1.05 \pm 0.02$ for the luminosity of large particles, $L_d$.

Calculations with other initial conditions produce similar results 
but on different timescales.  High velocity collisions convert 25\% 
or more of the initial mass into small particles `lost' to the grid.  
Cratering produces most of the mass loss; catastrophic disruption 
is responsible for $\lesssim$ 10\% of the lost mass. Collisional
damping reduces the velocities of the mid-sized particles before
cratering and catastrophic disruption convert all of the initial
mass into debris.  Because the collision lifetime of a particle 
scales with the particle density, collisional damping is more 
effective in the
more massive grids.  Once the eccentricities of the mid-sized
particles reach $\sim 5 \times 10^{-3}$, collisions begin to 
produce larger objects and less debris.  Dynamical friction then 
reduces the velocities of the largest bodies, while collisional
damping continues to `cool' the mid-sized and smaller bodies.
The most massive bodies eventually reach velocities small enough
to initiate runaway growth, where the largest bodies grow 
rapidly and the mass loss rate is negligible \citep{kl99}.
We plan to describe runaway growth and the formation of 
planets in a future paper.

Figure 3 outlines the evolution of a model with $x$ = 0.1 and 
$r_{max}$ = 10 km at $a_{mid}$ = 35 AU.  This model has fewer
small particles than a grid with $r_{max}$ = 10 m; the smaller 
collision rate thus leads to a smaller luminosity and a smaller 
damping rate.  The damping of the smallest particles is complete
by $t$ = 1 Myr, but the largest particles still retain most of
their initial eccentricities.  Some of the largest particles 
grow from accretion of small bodies, but most remain at their 
original masses (Fig. 3, middle panels).  Collisional damping 
continues until all of the debris-producing particles have 
small eccentricity at $t$ = 4 Myr.  Despite their large
eccentricities, the largest particles begin to grow rapidly
(Fig. 3, lower panels).  Dynamical friction then reduces their
velocities over the next 10 Myr.

To compare these and other results as a function of the starting
conditions, we derive the damping time $t_d$ from least-squares
fits to the luminosity evolution of each model.   We use the maximum 
dust luminosity $L_{max}$ for these fits.  Unless Poynting-Robertson
drag is efficient, the luminosity from small dust grains is larger 
than the luminosity contributed by larger objects in the main grid.  
The decline of the dust luminosity is therefore a convenient and 
relevant measure of the damping time.  In most models,  the dust
luminosity declines by a factor of $\sim$ 2 at $t = t_d$
and by a factor of $\sim$ 20 at $t = 10 t_d$.  

Figure 4 shows the variation of the damping time and maximum
dust luminosity with $x$ for models at 35, 70, and 140 AU. 
Models with the lowest surface densities are the least 
luminous and have the longest damping times.  In systems with
$x < 10^{-3}$, the damping time can exceed the stellar lifetime
of $\sim$ 1 Gyr.  For a given
surface density, calculations with larger $r_{max}$ have fewer
small particles.  These models therefore are less luminous
and have longer damping times.  Because the collision rate
declines with distance from the central star, the luminosity also
declines with increasing $a_{mid}$.  These results indicate that 
particle collisions in rings with $x > 10^{-2}$ can produce 
dust luminosities similar to those observed in a real debris 
disk system with a ring geometry.  Collisional damping reduces 
these luminosities
by an order of magnitude on timescales of 100 Myr or less.

To test the sensitivity of $L_d$ and $t_d$ to other initial
conditions, we vary the initial eccentricity $e_0$, 
the minimum and maximum particle size 
in the grid ($r_{min}$ and $r_{max}$);
the slope $\gamma$  of the initial mass distribution,
$N_C(m) \propto m^{-\gamma}$;
the fragmentation parameters $Q_c$ and $S_0$;
and the mass density of the particles $\rho_{ik}$
for calculations at $a_{mid}$ = 35, 70, and 140 AU.
Because the dust luminosity
is proportional to the collision rate, models with larger numbers
of small particles are more luminous than models with fewer small
particles.  These models also damp more rapidly. Thus, models with
$\gamma > 1$ and $r_{min} <$ 10 cm are more luminous and damp
more rapidly than models with $\gamma < 1$ and $r_{min} >$ 10 cm.
Models with `weaker' bodies -- smaller $Q_c$ and $S_0$ -- or with 
larger, less dense bodies are also more luminous and damp more rapidly. 
These models yield a simple relation between the damping time and
the luminosity of bodies in the grid at $t = 0$:
\begin{equation}
{\rm log} ~ ( t_d \cdot L_{d,0} / L_{\star} ) \approx 0.4 + 1.33 ~ {\rm log} ~ (a_{mid} / {\rm 35 ~ AU}) ~ .
\end{equation}
\noindent
The maximum dust luminosity at $t = 0$ is
\begin{equation} 
{\rm log} ~ ( t_d \cdot L_{max,0} / L_{\star} ) \approx 2.1 + 1.33 ~ {\rm log} ~ (a_{mid} / {\rm 35 ~ AU}) ~ . 
\end{equation}
\noindent
The scatter in the coefficients is $\pm$ 0.1 for a wide
range of initial conditions.
At late times, the dust luminosity follows a power law,
\begin{equation}
{\rm log} ~ L_{max} / L_{\star} \propto (t/t_d)^{-1.33 \pm 0.05} ~ .
\end{equation}
\noindent
For the luminosity evolution of the particles in the main grid, we derive 
\begin{equation}
{\rm log} ~ L_d / L_{\star} \propto (t/t_d)^{-1.0 \pm 0.1} ~ .
\end{equation}
\noindent

We identify only two exceptions to these relations.
In calculations with large $e_0$ or weak bodies, collisions can remove 
most of the initial mass on timescales shorter than the relevant 
damping time.  Figure 5 shows the ratio of final mass in the ring $M_f$
to the initial mass $M_i$ as a function of $e_0$ for planetesimals
with $S_0$ = $2 \times 10^6$ erg g$^{-1}$.  For modest perturbations
with $e_0 \lesssim 0.03$ $(a_i / {\rm 35 ~ AU})^{1/2}$, collisions
damp particle velocities before the ring loses a significant amount
of mass to small particles with $r_i < r_{min}$.  These models
follow the $L_d-t_d$ relations.  For strong perturbations with
$e_0 \gtrsim$ 0.04--0.06 $(a_i / {\rm 35 ~ AU})^{1/2}$, erosion
removes a significant amount of mass from the ring before damping
reduces particle velocities. These models have 25\%--50\% longer damping 
times; the asymptotic decline of the luminosity is identical to 
models with smaller perturbations, $L_d \propto t^{-1}$ for the 
large particles and $L_{max} \propto t^{-1.33}$ for the small particles.

In calculations with very large bodies ($r_{max} \gtrsim$ 500 km), 
viscous stirring and dynamical friction by the largest bodies in 
the grid can counteract collisional damping.  The dust luminosity 
can then remain large for timescales of $\sim$ 100 Myr or longer.  
Because collisional cascades in rings with embedded planets -- and 
no stellar flyby -- behave in a similar fashion, we plan to consider
these models in more detail in a separate paper.

\subsection{Cascades with Gas and Poynting-Robertson Drag}

Gas drag can be an important factor in planetesimal evolution 
\citep[e.g.,][]{nak83,wet93}.
Interactions between solid bodies and the gas damp particle velocities 
and remove particles from the model grid .  Both processes scale with 
the local gas density, $\rho_{g,i}$.  To understand how these processes 
affect the dust luminosity and the damping time,
we calculated a set of ring models for different initial values 
of the gas-to-dust mass ratio, $X_g \equiv \Sigma_g/\Sigma_d$.  
The interstellar medium has $X_g$ = 100; observed limits in debris
disk systems are uncertain \citep{zuc95,den95,lec98,gr00a,thi01}.
The appendix summarizes our treatment of particle removal and 
velocity damping.  We held the gas density fixed with time; 
$X_g$ therefore increases with time as fragmentation and gas drag 
remove solid bodies from the grid.  Calculations with $X_g$ = 1 are 
indistinguishable from models without gas.  Gas drag modifies the 
evolution of solid bodies for $X_g \gtrsim$ 10.  The following
paragraphs describe the evolution for $X_g$ = 10 and $X_g$ = 100.

Figure 6 shows the mass and velocity distributions at several times during 
the evolution of a model with $X_g$ = 10.  The initial conditions are 
identical to the model of Figure 1: small bodies with $r_{min}$ = 10 cm to 
10 m have the same initial eccentricity, $e_0$ = 0.02, and a total 
mass of 0.67 $M_{\oplus}$, in 16 annuli at $a_{mid}$ = 35 AU 
from a central 3 $M_{\odot}$ star.  Interactions between solid particles
and the gas have a modest impact on
the evolution of this model.  After 1 Myr, gas drag has swept less than 
0.01\% of the initial mass through the innermost annulus and out of the
grid.  Compared to a model without gas drag, orbital eccentricities are 
$\sim$ 5\% to 10\% smaller and the largest bodies are $\sim$ 10\% to 20\% 
larger (Figure 6, middle panels).  The amount of mass lost to small 
particles is $\sim$ 24\%, compared to $\sim$ 25\% in a model without 
gas drag.  After 3 Myr, gas drag is responsible for only $\sim$ 0.02\% 
of the total mass loss.  Orbital eccentricities are now $\sim$ 20\%
smaller, and the largest bodies are $\sim$ 40\% larger (Figure 6, 
lower panels).  This model reaches runaway growth earlier than a
model without gas drag; the final mass distributions of models
with and without gas drag are indistinguishable \citep{kl99}.

Interactions between solid bodies and the gas are much more important
when $X_g$ = 100 (Figure 7).  After 0.5 Myr, gas drag reduces
orbital eccentricities of the smallest bodies by a factor of four,
and the largest bodies begin to grow by mergers (Figure 7,
middle panels).  The mass lost from gas drag is small, $\sim$
0.1\% of the initial mass, compared to the mass lost from erosive
collisions, $\sim$ 22.5\%.  By 1 Myr, gas drag, dynamical friction,
and collisional damping cool the small bodies well before the
largest objects grow to sizes of 100 m (Figure 7, lower panels).
The small orbital eccentricities of this model yield an early 
runaway growth phase and the production of several Pluto-sized
objects on timescales of 10--30 Myr \citep{kl99}.  The mass of this
model at 1--3 Myr is $\sim$ 3\% larger than models without gas drag 
at similar phases; this difference does not yield significantly
larger objects after 10--30 Myr compared to models without gas drag.

Despite their more rapid evolution, models with gas drag follow nearly 
the same luminosity-damping time relation as models without gas drag.  
We derive log ($t_d \cdot L_{max,0}$) = $C + 1.33$ log ($a_{mid}$/35 AU)
for these models.  Damping is more efficient when gas drag is important;
larger gas-to-dust ratios imply smaller values of $C$.  Thus, the damping
times derived from models without gas drag are upper limits to the
lifetime of a collisional cascade produced by a stellar fly-by.  We 
also derive the same power-law decline of dust luminosity with time,
log $L_{max}/L_{\star} \propto (t/t_d)^{-1.33}$, as in models without
gas drag.

To complete this portion of our study, we consider mass loss from
Poynting-Robertson drag for various values of the stellar luminosity
and the particle mass density.
Mass loss from Poynting-Robertson drag is less than 1\% of losses 
from fragmentation and gas drag unless the stellar luminosity exceeds 
100 $L_{\odot}$ ($\rho_m/{\rm 1.5 ~ g ~ cm^{-3}}$). Damping of the 
horizontal velocity is also negligible.  For A-type stars with 
$L_{\star} \lesssim$ 100 $L_{\odot}$, Poynting-Robertson drag does not 
modify our conclusions unless the mass density of the mid-sized bodies
is much smaller than 1 g cm$^{-3}$.  Calculations with low density 
bodies yield smaller damping times than implied by equations
(1) and (2).  Although Poynting-Robertson drag generally is not 
important for the large grains considered here, removal of smaller
grains can modify the optical depth (see the appendix).  We plan 
to investigate this possibility in a future study.

\subsection{Cascades in Disks}

The ring calculations in \S3.1 and \S3.2 yield several interesting 
conclusions about collisional cascades in planetesimal disks.  
Stellar flybys can produce long-lived cascades in low mass disks, 
but the dust luminosity produced in these cascades is small,
$L_{max}/L_{\star} \simless 10^{-5}$ for $x \simless 10^{-2}$.
Cascades in rings with 1\% or more of the minimum mass solar nebula 
produce larger dust luminosities but have short observable lifetimes.  
The damping times in these models are $\sim$ 1--10 Myr.  These 
timescales are short compared to the ages of the oldest debris disk 
systems, $\sim$ 100--500 Myr \citep{hab99,so00a,spa01}.  

To test these conclusions in more detail, we consider more complete
disk models.  The disk consists of 64 annuli with $\Delta a_i/a_i$ 
= 0.025 and extends from $a_{in}$ = 30 AU to $a_{out}$ = 150 AU.
As in \S3.1, the disk contains bodies with $r_{min}$ = 10 cm to
$r_{max}$ and a mass density of 1.5 g cm$^{-3}$.  All of the bodies
have the same horizontal velocity with respect to a circular orbit;
the initial eccentricity in each annulus is $e_i$ = $e_1 a_i^{1/2}$
where $e_1$ is the eccentricity in the innermost annulus.  This 
condition yields a constant initial horizontal velocity in the disk.  
The mass spacing factor is $ \delta_1 = m_{k+1}/m_k$ = 2 for all 
batches; the initial mass distribution has equal mass per batch. 
We adopt the \citet{wet93} fragmentation algorithm with
$Q_c$ = $5 \times 10^7$ erg g$^{-1}$,
$S_0$ = $2 \times 10^6$ erg g$^{-1}$, and
$V_f$ = 1 m s$^{-1}$.

Figures 8--12 outline the evolution of a model with $x$ = 0.1, $r_{max}$ 
= 10 m, and $e_1$ = 0.02. The initial surface density is 
$\Sigma_i$ = 3 g cm$^{-2}$ $(a_i / {\rm 1 ~ AU} )^{-3/2}$;
the initial mass in solid material is 10 $M_{\oplus}$.
At the inner edge of the disk, collisions
rapidly damp orbital eccentricities on timescales of 1 Myr or less.
The largest bodies double their radius in the first 1 Myr and 
quadruple their radius in the first 10 Myr.  Because collision 
cross-sections scale with disk radius as $A \propto a_i^{-3}$, 
collisional damping proceeds very slowly at the outer edge of the disk.
Growth is slow.  Collisional damping starts to affect 
mid-sized 
bodies in the outer disk at 1 Myr, when bodies at $a_i \lesssim$ 40 AU 
are completely damped and starting to grow rapidly (Figure 8; 
middle panels).  Gravitational focusing leads to faster growth once 
mergers dominate erosive collisions.  Thus, large bodies at 30 AU
have entered the runaway growth phase when 10 cm bodies at 150 AU 
are just starting to grow (Figure 8; lower panels).

Figure 9 shows the sensitivity of collisional damping and of the growth 
rate of the largest objects to position in the disk.  The initial 
eccentricities of the smallest particles follow a power-law,
$e_i \propto a_i^{1/2}$; the scale height of these bodies above
the disk midplane is thus $H \propto a_i^{3/2}$.  The eccentricity 
of the smallest particles 
in the grid damps by a factor of $\sim$ 2 at $a_i$ = 30 AU in 1 Myr, 
when small particles at the outer edge of the grid have almost 
all of their initial orbital eccentricity.  At 10 Myr, damping has 
reduced the orbital eccentricities of all small particles by factors 
of 2 or more.  The eccentricities of the smallest particles then vary
with disk radius as $e_i \propto a_i^{0.85}$.  The power-law slope 
of this relation falls to $0.8$ at 50 Myr.  This behavior yields a 
very steep variation of the scale-height with radius, $H \propto a_i^{1.8}$.

The merger rate is more sensitive to disk radius.  For $t < 10$ Myr,
growth rates are linear, because gravitational focusing factors are small.
The radius of the largest body is a smooth function of disk radius,
$r_{i,max} \propto a_i^{-\gamma}$ with $\gamma \gtrsim$ 1.5.  Once particles
in the inner disk reach sizes of $\sim$ 1 km, gravitational focusing
factors increase and growth becomes non-linear.  Annuli in the
runaway growth phase depart from a smooth power-law variation of 
$r_{i,max}$ with $a_i$, as indicated by the sharp rise in $r_{i,max}$ 
for $a_i \lesssim$ 40 AU at 50 Myr. 

The total dust luminosity of this model follows the evolution 
derived for the ring models in \S3.1 (Figure 10).  At $t$ = 0,
the maximum dust luminosity of $L_{max}/L_{\star} \sim$ $6 \times 10^{-3}$
is comparable to luminosities observed in $\beta$ Pic and other luminous 
debris disk systems \citep{au99a,bar99,hab99,so00b}. The dust 
luminosity decreases 
with time and reaches $L_{max}/L_{\star} \sim$ $2 \times 10^{-5}$ at 
$t$ = 50 Myr.  This luminosity is close to luminosities observed in
$\alpha$ Lyr and other faint debris disk systems 
\citep{kal98,spa01}.  
Equation (1) describes the time-evolution of the luminosity remarkably 
well.  The luminosity-damping time relation is
\begin{equation}
{\rm log} ~ t_d \cdot L_{max,0}/L_{\star} = 3.16 \pm 0.05 .
\end{equation}
At late times, 
\begin{equation}
L_{max}/L_{\star} \approx 4 \times 10^{-3} ~ ( t/t_d ) ^{-1.03 \pm 0.02} ~ .
\end{equation}
\citet{kal98} derives $L \propto t^{-1}$ for known debris disk systems
with reliable ages.

To visualize the time evolution of the luminosity for this disk model, we 
assume a pole-on system where each annulus in the grid has a dust opacity 
derived in the geometric optics limit from the number density (\S A.6).  
The luminosity per unit surface area radiated by each annulus is then 
a simple function of the opacity and the dust scale height \citep{ken99}.  
We adopt the scale height for the smallest particles in the grid.
This procedure yields the bolometric surface brightness.  If the 
particle albedo is large and independent of grain size, the predicted 
surface brightness should be close to the optical or near-IR surface 
brightness of scattered light.  

Figure 11 shows how the radial surface brightness profile changes with 
time.  At the start of our calculations, models with a power-law surface 
density distribution $\Sigma_i$ produce a power-law surface brightness 
distribution $I_i$.  For $\Sigma_i
\propto a_i^{-3/2}$, we derive $I_i \propto a_i^{-7/2}$.  For 
$t \lesssim$ 1 Myr, collisions damp particle eccentricities
in the inner disk, $r_i \sim$ 30--60 AU.  Smaller collision velocities
produce less dust; the surface brightness fades.  At $t$ = 1 Myr,
the surface brightness is nearly constant for $r_i$ = 30--60 AU and
gradually approaches the initial $I_i \propto a_i^{-7/2}$ at
$r_i \gtrsim$ 100 AU.  As the calculation proceeds, collisions 
damp particle velocities in the outer disk.  After $t \approx$ 10 Myr,
the surface brightness in the inner disk declines by nearly two
orders of magnitude; the outer disk fades by $\sim$ 30\%. All of the
disk then fades by roughly an order of magnitude over the next 20--40
Myr.  When we terminate the calculation at 50 Myr, the surface brightness 
distribution is much more shallow than the initial distribution, with 
$I_i \propto a_i^{-1/2}$ instead of $I_i \propto a_i^{-7/2}$.

Figure 11 illustrates how coagulation concentrates mass into large
objects with small optical depth.  At 50 Myr, the radial gradient in
the surface density of solid material is similar to the initial gradient,
with $\Sigma \propto a_i^{-1.55}$.  Because material in the inner disk
has more frequent collisions than material in the outer disk, the inner
disk has more material in large objects with $r_i \gtrsim$ 100 m than
the outer disk. The surface density of small objects thus grows with
radius; for $t \gtrsim$ 10 Myr, $\Sigma_{small} \propto a_i^{1.5}$ for
$r_i \lesssim$ 10 cm.  There is some evidence for this behavior in
real systems: the particle disks in BD$+$31$^{\rm o}$643 \citep{kal97} 
and HD 141569 \citep{wei99} have regions where the number density of 
small particles appears to increase in radius.  We plan addititonal
calculations to see if this feature of the calculations is characteristic
of flyby models or all planet-forming particle disks.

Figure 12 shows disk images at six times in the evolutionary sequence.
An animation of the disk evolution is available in the electronic version 
of this paper.  To construct predicted images from the surface brightness 
profiles, we assume a simple power law relation to convert surface 
brightness to pixel intensity.  We simulate an observation by adding 
counting noise to the intensity of each pixel, but we do not convolve 
the image with a `seeing disk.' These approximations ignore radiative 
transfer effects in the disk and the variation of disk flux with 
wavelength, but the resulting images accurately illustrate the physical 
behavior of the dust luminosity with time.  

To investigate the sensitivity of these results to the starting conditions,
we consider several models with different values for $e_1$ and
$x$ and several models with gas drag.  Full disk models require more 
computer time than
ring models; this study is therefore more limited than the parameter
study of \S3.1.  As long as erosive collisions do not exhaust the supply
of mid-sized bodies before collisional damping can reduce particle
eccentricities, models without gas and Poynting-Robertson drag yield
the luminosity-damping time relation in equation (6) to an accuracy of
10\% or better independent of the initial conditions.  Low mass disks 
are less luminous and longer lived than massive disks.  Calculations 
with gas drag ($X_g$ = 100) evolve $\sim$ 30\% faster than models without 
gas drag.  Equation (5) thus provides an upper limit to the damping time 
for a disk with initial luminosity $L_{max,0}/L_{\star}$.

The evolution of the scale height is the main difference between models 
with and without gas drag.  When the gas density is low ($X_d \lesssim$
1--3), gas drag does not reduce significantly the orbital eccentricities 
of the smallest particles in the grid.  Collisional damping and dynamical
friction dominate the velocity evolution; the scale height of small 
particles exceeds the scale height of large particles.  When the
gas density is larger ($X_d \gtrsim$ 10), the small particles are more
closely coupled to the gas and have more circular orbits.  The mid-sized
particles then have the largest scale heights.  If the gas and small 
particles are well-mixed perpendicular to the disk midplane, the dust 
has a large scale height and a large dust luminosity \citep{kh87}.
If mixing between dust and gas is poor, small dust particles settle 
to the midplane; the dust luminosity is then small.  Detailed
hydrodynamical calculations are necessary to derive the outcome of a
collisional cascade in a planetesimal disk with large amounts of gas.

Full disk calculations yield the same behavior with initial eccentricity
as ring models.  For annuli with $e_0 \lesssim 0.05$ 
$(a_i / {\rm 35 ~ AU})^{1/2}$, collisional damping reduces particle 
velocities before the disk loses more than half of its initial mass.  
The disk loses more than 95\% of its initial mass for strong perturbations 
with $e_0 \gtrsim 0.05$ $(a_i / {\rm 35 ~ AU})^{1/2}$.  In these strong
perturbations, collisional damping reduces particle velocities sufficiently 
to allow growth by mergers and the formation of 1 km or larger bodies.  
The timescale for planet formation is then very long, $\sim$ 1 Gyr 
or longer.

Our results indicate that complete disk models evolve on faster timescales
than ring models \citep[see also][]{kb01}.  
In ring models, planetesimals at the edges of the grid 
interact with fewer annuli than planetesimals in the middle of the
grid.  Collisional damping is less effective with fewer collisions. Partial 
disk models thus damp more slowly and lose more mass than complete disk models.
At 35 AU, complete disk models lose $\sim$ 20\% less mass than ring
models.  This difference rises to $\sim$ 50\% at 140 AU.  More effective
damping and less mass loss yield larger bodies on shorter timescales 
in complete disk models.  

\subsection{Limitations of the Models}

Statistical simulations of planetesimal growth in a dusty disk have a 
long history, with well-known limitations and uncertainties 
\citep{wet80,gre84,dav85,bar91,spa91,lis93,wet93,st97a,wei97,kl98,ina01}.
The particle-in-a-box formalism assumes a homogeneous spatial distribution
of solid bodies with small velocities relative to the local circular 
velocity.  This assumption is good until runaway growth produces a 
few massive bodies not distributed uniformly in space \citep{kok96}.  
Our simulations never reach this limit.  Our calculations also 
require a specific algorithm for collision outcomes (\S2 and the appendix).  
As long as particle collisions produce damping and some debris, our
derived relations for the evolution of luminosity as a function of 
initial disk mass are remarkably independent of the details of the
collision algorithm.  The collisional damping prescription is more
important; changing the coefficients in the algorithm (equations
A8 and A9 of the appendix) changes the damping times by a similar
factor.  We have not investigated the sensitivity of the results to
the form of the damping equations.  We have tested our prescription 
for velocity evolution in the low velocity limit, where the 
relative collision velocity is small compared to the Hill velocity
\citep{bar90,wet93,kl98,ste00,kb01}.  The 
statistical approach is invalid in this limit. \citet{ida90}, 
\citet{bar90}, \citet[][1993]{ida92}, and \citet{wet93} have described
solutions to this problem; we use the \citet{ida92} approach.
Tests with other approaches yield the same results.

The implementation of the algorithms produces other limitations.  Our 
multiannulus calculations eliminate many of the constraints associated 
with calculations of a single accumulation zone 
\citep{wet90b,wet93,wei97,kl99}.  We can treat
collisions and gravitational perturbations between particles in adjacent 
annuli, and the drift of particles through adjacent annuli due to gas drag 
and Poynting-Robertson drag.  The current calculations do not include 
algorithms for orbital migration of planetesimals or the evolution of 
gas in the disk
(including gas accretion), but these processes are not important for
the situations we investigate here.  

Uncertainties resulting from the finite mass and spatial resolution 
should be small.  Calculations with coarse mass resolution within 
an annulus lag higher resolution simulations by 10\% to 20\% but 
yield nearly the same mass and velocity distributions 
\citep[see][1999]{kl98}.  
Calculations with finer spatial resolution are indistinguishable from
the simulations described above at the 5\% level.  These differences 
grow with the mass of the largest object in the grid but are not 
important for the relatively small mass objects we consider here.

To compute the dust luminosity, we assume two limits to the dust 
opacity of small objects.  We make these assumptions to avoid computing 
the time evolution of small particles, which have a 
negligible impact on the evolution of larger planetesimals in the grid.  
In most planetesimal disks, the large opacity limit is appropriate, 
because the timescales for Poynting-Robertson drag are $\sim$ 1 Myr
or longer at 30 AU and $\sim$ 25 Myr or longer at 150 AU.  These 
timescales are long compared to the damping times of massive, luminous
planetesimal
disks.  For more luminous central stars with $L_{\star} \sim 100 ~ L_{\odot}$, 
the small opacity limit is appropriate, because Poynting-Robertson drag 
can remove small grains on timescales much shorter than 1 Myr at 30 AU.
We plan to address these uncertainties in more detail in a future paper.

In our complete disk models, we adopt a shallow variation of the initial
eccentricity of planetesimals with disk radius, $e_0 \propto a_i^{1/2}$.
The ratio of the initial horizontal velocity to the shattering velocity
of a dust grain is then independent of $a_i$, which allows us to 
investigate collisional damping without worrying about the importance
of different fragmentation algorithms.  This approximation is reasonable 
for moderately strong perturbations where the passing star has a 
periastron of $\sim$ 600 AU.  Closer stellar flybys produce stronger 
perturbations with $e_0 \propto a_i^{5/2}$ \citep{ida00,kal00,kob01}.
In a strong perturbation, planetesimals in the outer disk are depleted
more rapidly than those in the inner disk.  Even if the outer disk loses
nearly all of its initial mass, our calculations indicate that collisional 
damping should reduce particle velocities and allow planet formation on
long timescales. We plan to investigate this possibility in a future paper.

The speed of the computer is a final limitation to these calculations.
Although larger test simulations are possible, practical considerations 
limit calculations of model grids to 64--128 annuli.  For reasonable
spatial resolution, the maximum extent of our grid is a factor of 5--10
in distance from the central star.  We are thus 1--2 orders of magnitude
from constructing model grids of complete solar systems.  For a given
set of algorithms, the timescales and dust luminosities derived here should
be within 10\% to 50\% of those derived from more extensive models.
Limitations on the extent of the radial grid are more important for
addressing the interfaces between (i) gas giants and terrestrial planets 
and (ii) gas giants and the Kuiper Belt in our solar system.  Similar
problems occur in evolutionary calculations of other types of 
circumstellar disks.  Faster computers should resolve these 
difficulties in the next few years.

\section{DISCUSSION AND SUMMARY}

Stellar flybys can plausibly produce collisional cascades in a 
planetesimal disk.  When the mass of solid material in the disk 
is 1\% or more of the minimum mass solar nebula 
($\sim$ 1--100 $M_{\oplus}$ of solid material at 30--150 AU),
flybys which increase planetesimal velocities close to the shattering 
limit can produce dust luminosities similar to those observed in 
known debris disk systems.  The dust luminosity correlates with the 
disk mass.  Collisional damping in these disks is very efficient; 
dust production decreases dramatically on timescales of 1 Myr or less.  
The most luminous and most massive disks damp the fastest.  These 
conclusions are independent of many uncertainties in the calculations, 
including the initial mass distribution, the fragmentation parameters, 
and the mass density of planetesimals.  

Our calculations indicate two outcomes for a collisional cascade induced 
by a moderately close stellar flyby.  If the flyby produces relative 
velocities which exceed the strength of a planetesimal, disruptive 
collisions nearly exhaust the supply of planetesimals before collisional 
damping becomes effective.  Most planetesimals are eventually ground to 
dust and planet formation is difficult \citep[see also][]{kob01}.  
For icy planetesimals with 
$S_0 = 2 \times 10^6$ erg g$^{-1}$, this limit is $e_0 \gtrsim$ 
0.05 $(a_i / {\rm 35 ~ AU})^{1/2}$.

When a flyby produces a smaller perturbation in the disk, collisional 
damping reduces planetesimal velocities on relatively short timescales.  
This process allows the disk to retain a significant mass in large 
planetesimals, which collide and merge to form planets.  Our ability to 
detect the collisional cascade observationally depends on the disk mass. 
Collisional cascades in disks with $\sim$ 1\% or more of the minimum mass 
solar nebula produce dust luminosities, $L_{max}/L_{\star} \gtrsim 10^{-5}$,
comparable to those observed in debris disk systems \citep{kal98,spa01}.

Collisional damping is a new feature in models of the evolution of collisional
cascades in planetesimal disks.  Previous models assumed that collisions
preserve the eccentricity and inclination distributions of planetesimals
in the disk \citep[e.g.,][see also Artymowicz 1997, Lagrange et al. 2000,
and references therein]{mou97,aug01}.  Dust 
grains produced by collisions thus had the same orbits as their progenitors.
In our models, `collisionless' systems with long collision timescales
yield optical depths and dust luminosities much smaller than those 
observed (Figures 4 and 10).  Systems with larger optical depths and dust 
luminosities have short collision times and short damping timescales. 
This conclusion is an extension of earlier work on the debris disk system
HR 4796A and the Kuiper Belt in our solar system, where disks with 
masses of solid material comparable to or larger than 
the minimum mass solar nebula allow the growth of Pluto-sized objects,
which stir the velocities of smaller objects to the shattering 
velocity and produce observable collisional cascades \cite[][see 
also Stern \& Colwell 1997a,b]{kl99,ken99}.

Following a flyby, equation (1) fits the evolution of the dust luminosity 
to remarkably high accuracy.  Because damping and dust production depend on
the collision rate, collision physics sets the form of this relation.  To
derive the limiting form of equation (1), we assume an ensemble of identical 
planetesimals.  The coagulation equation is $ dn/dt \propto n^2 $ (\S A.2).
The dust production rate is $dm/dt \propto m^2 n^2 \propto n^2$ where $m$ 
is the mass of a planetesimal.  Thus, the disk mass and the number density 
of planetesimals fall with time, $M_d \propto t^{-1}$.  Because the disk
has constant surface area, the surface density follows the same relation,
$\Sigma \propto t^{-1}$.  The left panels of
Figure 3 verify this relation for the damping time, 
$t_d \propto \Sigma_d^{-1}$.  The luminosity of the planetesimals is 
proportional to the disk mass, which yields $L \propto t^{-1}$; our models
follow this relation faithfully (Figure 2 and 10).  The dust luminosity
depends on the mass distribution of planetesimals for $r_k < r_{min}$,
where $r_{min}$ is the minimum planetesimal radius in the grid.  The
appendix describes two limits on this mass distribution. The maximum
dust mass assumes that Poynting-Robertson and other drag forces do not remove
particles from the grid; the minimum dust mass assumes that drag forces
remove particles efficiently from the grid.  In both limits, the dust
mass is proportional to the disk mass (Figure 3, right panels).  The 
dust luminosity thus declines with time, 
\begin{equation}
L_{max,min} \propto t^{-1}.
\end{equation}
\noindent
\citet{kal98} derives the same relation from observations of debris disks.
Our detailed models for complete disks yield this relation; because
planetesimals have fewer annuli to interact with, the relation for 
ring models is steeper.

\citet{spa01} derive a different relation for the time-variation of
luminosity in a collisional cascade, $L_d \propto t^{-2}$.  
They assume that the dust clearing time is shorter than the age of the 
system and conclude that the dust luminosity is proportional to the 
instantaneous dust production rate, $dn/dt$.  However, the dust 
luminosity is proportional to $dn/dt$ only if planetesimal collisions 
produce dust which escapes the system on timescales shorter than the 
collision time.  Because Poynting-Robertson drag and gas drag are slow 
processes in the outer disk, radiation pressure is the only mechanism 
capable of removing particles from a large debris disk on short 
timescales\footnote{Poynting-Robertson and gas drag can efficiently
remove small particles from a planetesimal disk at distances of 
1--5 AU from the central star.  Relatively rapid dust removal might
account for the large inner `holes' observed in some debris disk
systems \citep[see also][]{bac93,art97,koe98,lag00}.}.
Particles with sizes of 10 $\mu$m or larger are unaffected by radiation
pressure.  Planetesimal collisions thus build up a reservoir of dust
grains with sizes of 10 $\mu$m or larger, while radiation pressure 
ejects smaller grains \citep{bur79,bac93,art97}.  For the power law
cumulative mass distribution appropriate for collision debris, 
$N_C(m) \propto m^{-0.83}$ \citep{doh69,dav85,wet93,kl99}, the optical depth
of grains ejected by radiation pressure is a factor of 10 or more 
smaller than the optical depth of larger grains.  The larger grains
in an annulus thus make a larger contribution to the dust luminosity
than smaller grains accelerated away from the annulus.  The decline
of the dust luminosity then depends on the rate of decline for the
total disk mass, $L_{max} \propto M_d \propto t^{-1}$.

Our results suggest modifications of the picture for the formation and 
early evolution of the Kuiper Belt.  \citet{ida00} propose a stellar 
flyby to explain
the structure of the outer Kuiper Belt in our solar system.  In their model,
a stellar encounter with a distance of closest approach of $\sim$ 100 AU
excites the large orbital eccentricities and inclinations observed for
50--500 km objects at distances of 30--50 AU from the Sun.  Collisions
between these objects lead to a collisional cascade which erodes the mass 
of the inner Kuiper Belt over time.  In our calculations, collisional damping 
of 1--100 m planetesimals accompanies mass erosion; dynamical friction 
then damps the eccentricities and inclinations of the largest objects on 
short timescales, $\sim$ 1--10 Myr for a minimum mass solar nebula model
and $\sim$ 100 Myr for a disk with a mass of solid material equal to 
the mass of the current Kuiper Belt.  
Our models predict damping times of 5 Gyr or longer only when the disk has 
a small mass in solid objects with radii of 1 km or less.  Coagulation models 
which can produce 50--500 km Kuiper Belt objects at 30--40 AU leave most 
of the initial mass in smaller objects with radii of 0.1--10 km 
\citep[][see also Stern \& Colwell 1997a,b]{ken02,kl99}.
Thus, if a stellar encounter is responsible for the large orbital
eccentricities observed in Kuiper Belt objects, our models suggest that
this encounter occurred after some other process depleted the Kuiper Belt
of 0.1--1 km objects.  We plan to investigate the idea that stirring 
by one or more Pluto-sized objects embedded in the Kuiper Belt can 
explain this depletion.

\citet[][see also Kalas et al. 2000; Kalas, Deltorn, \& Larwood 2001; 
Larwood \& Kalas 2001]{kal95} propose a stellar flyby model to 
explain structures observed in the $\beta$ Pic debris disk.  In their 
picture, a recent, $\sim 10^5$ yr, stellar encounter produces an 
asymmetric ring system and other dynamical features in the disk 
\citep{kal00}.  The highly perturbed orbits of planetesimals in
the disk lead to a collisional cascade and copious dust production.  
Although we cannot calculate planetesimal evolution in the $\sim$ 
1000 AU $\beta$ Pic disk with adequate spatial resolution, our 30--150 AU 
disk model in \S 3.3 provides a reasonable first approximation to the 
evolution of a collisional cascade induced by a moderately close 
stellar flyby in the inner portion of a 
large disk.  The maximum dust luminosity of $L_{max,0} / L_{\star} \sim$ 
$4 \times 10^{-3}$ is close to the bolometric luminosity ratio $L_b 
/ L_{\star} \sim$ $3 \times 10^{-3}$ observed in $\beta$ Pic \citep{bac93}.  
The damping time of the model, $\sim 3 \times 10^5$ yr, indicates that
collisions do not wash out either the observed ring structure or the radial 
variation of eccentricity and inclination at distances of 100--1000 AU
on timescales similar to the apparent dynamical age of $\sim 10^5$ yr.   
However, our models produce significant collisional damping at 30--50 AU 
on short timescales.  Damping produces an apparent hole in the disk, with 
smaller dust production than annuli in the outer disk (Figures 11--12).
In principle, our models can produce changes in the slope of the radial 
surface brightness distribution similar to that observed in the $\beta$ 
Pic disk \citep{gol93,aug01}.  
Radiative transfer calculations similar to those of 
\citet{ken99} will allow us to make a detailed comparison between our 
models and observations.

Associating stellar flybys with other debris disk systems seems less
practical.  The damping time-luminosity relation (equation (6)) implies 
that all collisional cascades induced by stellar flybys have luminosities
which decline below $L_{max}/L_{\star} \sim 10^{-5}$ on timescales 
of 100 Myr or less.  Source statistics suggests that nearly all stars 
are born with a disk \citep{lad99}; many main sequence stars also
have disks \citep{hab01}. Several known debris disk systems have 
ages of 500 Myr or more \citep{so00b,spa01}. Recent flybys for all
objects are unlikely; the chance probability of a stellar encounter 
which passes within $\sim$ 500 AU of a nearby field star is 1\% or 
less in $\sim$ 100 Myr \citep{fro98,gar99}.  We suspect that stirring 
by embedded planets or a binary companion induces collisional cascades 
in most debris disk systems.  We plan future papers to address this issue.

Our results suggest caution when interpreting observed relations between 
dust luminosity and age in debris disk systems.  Most evolutionary
processes in a planetesimal disk scale with the orbital period. The factor 
of 4--6 range in the stellar mass among known debris disk systems implies 
a range of 2--2.5
in evolutionary timescales for systems with the same initial disk mass
and disk radius.
In an ensemble of stars with a range of temperatures and luminosities, 
Poynting-Robertson drag will cause lower dust luminosities in disks with 
luminous stars compared to less luminous stars.  Poynting-Robertson and 
gas drag may also reduce disk luminosities below current detection limits 
on shorter timescales for luminous stars than for less luminous stars.  
This behavior suggests that the timescale for disk evolution should correlate
with the nature of the central star. \citet{so00b} comment on this effect
in their data for 10--15 debris disk systems.  Although the trend described
by \citet{so00b} is encouraging, current samples are too small and the
selection effects too uncertain to make robust statements about timescales
for disk evolution.  Larger samples with smaller detection limits may
provide enough data to test trends predicted by the models in detail.

Finally, multiannulus accretion codes are an important step towards 
understanding the formation and evolution of planetary systems
\citep[see also][]{spa91,wei97}.  Our calculations demonstrate that 
`complete' disk models with 64 or more annuli evolve on faster timescales
than ring models with 16 or fewer annuli.  Planetesimals at the 
edge of a model grid interact with fewer annuli than planetesimals in 
the middle of a grid.  Collisional damping, dynamical friction, and 
runaway growth thus proceed more rapidly in a complete disk model than 
in a ring model.  Calculations of fragmentation, gas drag, and 
Poynting-Robertson drag are also more accurate in a multiannulus grid.
This difference is important for constructing accurate models for the
formation of Jupiter and the Kuiper Belt, where observational constraints
limit the mass and timescale available for planet formation 
\citep[e.g.,][]{lis87,bai94,pol96,kl99,bry00}

\vskip 4ex

We acknowledge a generous allotment, $\sim$ 600 cpu days, of 
computer time on the Silicon 
Graphics Origin-2000 `Alhena' through funding from the NASA Offices 
of Mission to Planet Earth, Aeronautics, and Space Science.  Advice 
and comments from M. Geller, M. Kuchner, and an anonymous referee 
greatly improved our presentation.  

\vfill
\eject

\appendix

\section{APPENDIX}

\subsection{Overview}

We assume that planetesimals are a statistical ensemble of masses 
in $N$ concentric, cylindrical annuli with width $\Delta a_i$
and height $H_i$ centered at radii $a_i$ from a star with mass
$M_{\star}$ and luminosity $L_{\star}$.  
The particles have horizontal $h_{ik}(t)$ and vertical $v_{ik}(t)$
velocity dispersions relative to an orbit with mean Keplerian velocity 
$V_{Ki}$ \citep{lis93,ste00}.  We approximate the continuous distribution 
of particle masses with discrete batches having an integral number of 
particles $n_{ik}(t)$ and total masses $M_{ik}(t)$.  The average mass 
of each of $M$ mass batches, $m_{ik}(t)$ = $M_{ik}(t) / n_k(t)$, evolves 
with time as collisions add and remove bodies from the batch \citep{wet93}.  

\citet[][1999]{kl98}
describe our approach for solving the evolution of particle numbers 
and velocities for a mass batch $k$ in a single annulus $i$ during 
a time step $\delta t$.  \citet{kb01} describe our solution for the 
velocity evolution from elastic collisions for a set of annuli.  Here 
we describe collision rates and velocity evolution from inelastic 
collisions for a set of annuli. We also include our treatment of 
particle loss and velocity evolution for gas drag and the 
Poynting-Robertson effect.

\subsection{Coagulation Equation}

The coagulation equations for a particle in mass batch $k$ of annulus $i$
colliding with another particle in mass batch $l$ of annulus $j$ are

\begin{equation}
\delta n_{i^{\prime}k^{\prime}} = \delta t \left [ \epsilon_{ijkl} A_{ijkl} n_{ik} n_{jl} 
~ - ~ n_{i^{\prime}k^{\prime}} A_{i^{\prime}jk^{\prime}l} n_{jl} \right ] ~ + ~ \delta n_{i^{\prime}k^{\prime},f}~ - ~ \delta n_{i^{\prime}k^{\prime},gd} ~ - ~ \delta n_{i^{\prime}k^{\prime},prd}
\end{equation}

\begin{equation}
\delta M_{i^{\prime}k^{\prime}} = \delta t ~ m_{i^{\prime}k^{\prime}} \left [ \epsilon_{ijkl} A_{ijkl} n_{ik} n_{jl} ~ - ~ n_{i^{\prime}k^{\prime}} A_{i^{\prime}jk^{\prime}l} n_{jl} \right ] ~ + ~ \delta m_{i^{\prime}k^{\prime},f} - ~ \delta m_{i^{\prime}k^{\prime},gd} - ~ \delta m_{i^{\prime}k^{\prime},prd}
\end{equation}

\noindent
where $A_{ijkl}$ is the cross-section,
$\epsilon_{ijkl} = 1/2$ for $i = j$ and $k = l$, and 
$\epsilon_{ijkl} = 1$ for $k \ne l$ and any $i, j$.
The terms in A1--A2 represent (i) mergers of $m_{ik}$ and $m_{jl}$ 
into a body of mass $m_{i^{\prime}k^{\prime}} = m_{ik} + m_{jl} - m_{e,ijkl}$, 
(ii) loss of $m_{i^{\prime}k^{\prime}}$ through mergers with other bodies, 
(iii) addition of $m_{i^{\prime}k^{\prime}}$ from debris produced
by the collisions of other bodies, and (iv) loss of $m_{i^{\prime}k^{\prime}}$ 
by gas drag and Poynting-Robertson drag.  \citet{kl99} outlines our calculation
of the mass lost to small fragments, $m_{e,ijkl}$.  The second term 
in A1--A2 includes the possibility that a collision can produce debris 
but no merger \citep[rebounds; see][and references therein]{dav85,bar93,kl99}.

Following Weidenschilling et al. (1997), particles in adjacent annuli 
may collide if their orbits approach within 2.4 times their mutual Hill 
radius $R_H$. The `overlap region' for these inelastic collisions is 

\begin{equation}
o_{ijkl,in} = 2.4 R_H + 0.5 (w_{ik} + w_{jl}) - | a_i - a_j | ~ ;
\end{equation}

\noindent
where $w_{ik}$ is the radial extent of the orbit of particle
$k$ with orbital eccentricity $e_k$ in annulus $i$ \citep{kb01}:

\begin{equation}
w_{ik} = \left\{ \begin{array}{l l l}
         \Delta a_i + e_k a_i & \hspace{5mm} & e_k a_i \le \Delta a_i \\
         (\Delta a_i + e_k a_i) (e_k a_i / \Delta a_i)^{1/4} & \hspace{5mm} & e_k a_i > \Delta a_i \\
 \end{array}
         \right .
\end{equation}

\noindent
We calculate the `gravitational range' of the largest bodies --
$R_{g,ik} = K_1 a R_{H,ijkl} + 2 a e_{ik}$ \citep{wet93} --
where $K_1 = 2 \sqrt{3}$ and $R_{H,ijkl} = [(m_{ik} + m_{jl})/3 \msun]^{1/3}$
is the mutual Hill radius \citep[see also][1999]{kl98}.  Isolated bodies 
are the $N$ largest bodies that satisfy the summation,
$ \sum_{k_{min}}^{k_{max}} ~ n_{ik} R_{g,ik} \ge \Delta a_i$ \citep{wet93}.
These isolated bodies cannot collide with one another but can
collide with other lower mass bodies in the annulus and all other
bodies in other annuli.

We use the particle-in-a-box technique to calculate collision cross-sections,
$A_{ijkl}$.  \citet{kl98} describe this approach for a calculation with a single 
annulus.  The technique is easily generalized to a multi-annulus calculation
by averaging quantities which depend on the radius of the annulus, such
as the Keplerian velocity or the Hill radius \citep{spa91,wei97}.  
The cross-section is

\begin{equation}
A_{ijkl} = \alpha_{coll}~\left ( \frac{1}{4~H_{ijkl}~\langle a_{ij} \rangle~ \langle \Delta a_{ij} \rangle} \right ) ~V_{ijkl}~F_{g,ijkl}~(r_{ik} + r_{jl})^2 ~ ,
\end{equation}

\noindent
where 
$\alpha_{coll}$ is a constant \citep{wet93, kl98},
$H_{ijkl}$ is the mutual scale height,
$\langle a_{ij} \rangle $ and $ \langle \Delta a_{ij} \rangle $ are the 
average heliocentric distance and width for the two interacting annuli,
$V_{ijkl}$ is the relative particle velocity,
$F_{g,ijkl}$ is the gravitational focusing factor, and
$r_{ik}$ and $r_{jl}$ are the particle radii.
We adopt the piecewise analytic approximation of \citet{spa91} 
for the gravitational focusing factor, and the two body collisional
cross-sections of \citet{gre91} in the low velocity limit
\citep[see also][1992]{grz90}.

To calculate $i^{\prime}$ and $k^\prime$ for each collision,
we establish fixed grids of annuli and masses.  The radial grid has
annuli with either a fixed constant size,
$\Delta a_i = a_{i+1} - a_i$, or a variable width,
$\Delta a_i = \Delta a_0 \cdot a_i$.
The mass grid in each annulus has particle numbers
$k = 1$ to $k = N_{max}$ with $\delta_k$ = $m_{k+1}/m_k$.
We set
\begin{equation}
\delta_k = \left\{ \begin{array}{l l l}
        \delta_1& \hspace{5mm} & k \le k_0 \\
        \delta_1 - \epsilon \cdot (k - k_0) & \hspace{5mm} & k_0 < k < k_1 \\
        \delta_2 & \hspace{5mm} & k \ge k_1 \\
        \end{array}
        \right .
\end{equation}
\noindent
where $\delta_1$, $\delta_2$, and $\epsilon$ are constants.
Grids with a fixed mass ratio between neighboring mass bins 
have $\epsilon \equiv$ 0.  Grids with $\epsilon >$ 0 allow coarse
resolution, $\delta \approx$ 1.4--2.0, where small objects grow slowly,
and fine resolution, $\delta \approx$ 1.05--1.4, where large objects
grow rapidly. When a collision produces $n_{k^\prime}$ bodies with 
$m_{k^\prime}$ in annulus $i^\prime$, we augment either batch $l^\prime$ 
when $m_{k^\prime} \le \sqrt{m_{l^\prime} m_{l^\prime+1}}$ or batch 
$l^\prime+1$ when $m_{k^\prime} > \sqrt{m_{l^\prime} m_{l^\prime+1}}$.  
We add mass to annulus $j^\prime$
when $f - j^\prime < 0.5$ and to annulus $j^\prime+1$ when
$f - j^\prime \ge 0.5$, where
\begin{equation}
f = \frac{i m_{ik} + j m_{jl}}{m_{ik} ~ m_{jl}} ~ ,
\end{equation}
\noindent
and $j^\prime$ is the integer part of $f$.  A complete cycle through 
all mass batches and all annuli produces new values for $n_{k^\prime}$ 
and $M_{k^\prime}$ in each annulus $i^\prime$, which yields new values 
for the average mass per bin, $m_{k^\prime} = M_{k^\prime}/n_{k^\prime}$.
This process conserves mass and provides a good description of 
coagulation when $\delta_k$ is small \citep[e.g.,]{oht88,wet90a,oht90,kl98}.
Our algorithm to place merged objects in a particular annulus does not
conserve angular momentum specifically; test calculations conserve
angular momentum to better than 1\% after $10^6$ timesteps.

\subsection{Velocity Evolution}

The time evolution of particle velocities from collisional damping is
\begin{equation}
\frac{dh_{in,ik}^2}{dt} = \sum_{j=0}^{j=N} \sum_{l=0}^{l=l_{max}} \frac{C_{in}}{2}~(m_{jl} h_{jl}^2 - m_{ik} h_{ik}^2 - (m_{ik} + m_{jl}) h_{ik}^2)~I_e
\end{equation}

\begin{equation}
\frac{dv_{in,ik}^2}{dt} = \sum_{j=0}^{j=N} \sum_{l=0}^{l=l_{max}} \frac{C_{in}}{\beta_{ijkl}^2}~(m_{jl} v_{jl}^2 - m_{ik} v_{ik}^2 - (m_{ik} + m_{jl}) v_{ik}^2)~I_i
\end{equation}
\noindent
where 
$C_{in} = \alpha_{coll} ~ f_{ijkl} ~ \epsilon_{ijkl} ~ \rho_{g,jl} ~ 
V_{ijkl} ~ F_{g,ijkl} ~ (r_{ik} + r_{jl})^2$ \citep[][1999]{hor85,wet93,oht92}.
In the second summation, $l_{max} = k$ when $i = j$;
$l_{max}$ = $M$ when $i \neq j$ \citep[see also][1999]{kl98}.  
We add a term, $f_{ijkl}$, to treat the overlap between adjacent
zones; $f_{ijkl}$ = 1 when $i = j$ and $f_{ijkl} \leq 1$ when
$i \neq$ j.  The integrals $I_e$ and $I_i$ are elliptic integrals
described in previous publications 
\citep[][1999, Stewart \& Ida 2000]{hor85,wet93,oht92}.

\subsection{Gas Drag}

Gas drag can be an important process during the early evolution of
planetesimal disks.  The viscosity of the gas drags small bodies
inwards through the disk and damps the motion of small bodies 
relative to a circular orbit.  \citet{ada76} and \citet{wei77b} analyze 
the interactions between gas and planetesimals, and develop useful
evolution formulae \citep[see also][]{kar93}.  
We adopt the \citet{ada76} prescription.
Material drifts through an annulus at a rate
\begin{equation}
\frac{\delta a_{ik}}{a_i} = 2 (0.97 e_{ik} + 0.64 i_{ik} + \eta_{ik}/V_{K,i})
                    \frac{\eta_{i}}{V_{K,i}} ~
                    \frac{\delta t}{t_{d,ik}} ~ ,
\end{equation}
\noindent
where $\eta_{i}$ is the gas velocity relative to the local Keplerian
velocity, $V_{K,i}$.  The characteristic drift time is

\begin{equation}
t_{d,ik} = \frac{365}{C_D}~\left ( \frac{m_{ik}}{\rm 10^{21}~g} \right )^{1/3}
         \left ( \frac{\rm 1~AU}{a_i} \right )
         \left ( \frac{10^{-9}~\rm g~cm^{-3}}{\rho_{g,i}} \right ) T_{K,i},
\end{equation}

\noindent
where $C_D$ = 0.5 is the drag coefficient, $\rho_{g,i}$ is the gas density,
and $T_{K,i}$ is the orbital period \citep{ada76,wet93}.
The removal of bodies from annulus $i$ at each time step is then
\begin{equation}
\frac{\delta n_{gd,ik}}{n_{ik}} = \frac{\delta a_{ik}}{a_i} ~ .
\end{equation}

To specify the gas density, we adopt an initial gas-to-dust surface
density ratio, $\Sigma_{g0,i} / \Sigma_{d0,i}$, and set 
$\rho_{g0,i} = \Sigma_{g0,i} / 2 H_{g0,i}$, where $H_{g0,i}$ 
is the vertical scale height of the gas.  We assume a scale height
appropriate for a gaseous disk in hydrostatic equilibrium:
\begin{equation}
H_{g0,i} = H_0 (a_i/a_0)^{q} a_i ~ ,
\end{equation}
\noindent
where $H_0$ is the scale height at $a_0$ = 1 AU. In most cases,
we adopt $H_0$ = 0.05 and $q = 0.125$ \citep{kh87}.  We also assume
that the gas density falls with time:
\begin{equation}
\rho_{g,i} (t) = \rho_{g0,i}~e^{-t/t_g}
\end{equation}
\noindent
where $t_g$ is a constant and is usually 10 Myr to 1 Gyr.

Our gas drag algorithm removes bodies from annulus $i+1$ and places
them in annulus $i$.  We explicitly conserve kinetic energy: the
velocity dispersion of annulus $i$ decreases as lower-velocity 
material is dragged inwards from outer annuli. The grid loses 
material and kinetic energy at the inner boundary and gains 
material and kinetic energy at the outer boundary.   
To compute the amount of material added to the grid,
we assume a power-law density distribution, $\rho_{g,i} \propto
a_i^{-\alpha}$.  Equation (A12) is then $\delta n_{gd,ik} / n_{ik}$
$\propto a_i^{-(\alpha+0.5)}$, where we have ignored the slow 
variation of $e$, $i$, and $\eta_i/V_{K,i}$ with $a_i$.  The number
of particles added to the last zone in the grid is:
\begin{equation}
\delta n_{gd,Nk}^{\prime} = \left ( \frac{a_{N}}{a_{N+1}} \right )^{\alpha+0.5} ~ \delta n_{gd,Nk} ~ .
\end{equation}
\noindent
The addition of kinetic energy follows
\begin{equation}
\delta E_{gd,Nk}^{\prime} = \left ( \frac{a_{Nk}}{a_{N+1k}} \right )^{\alpha+1.5} ~ \delta E_{gd,Nk} ~ ,
\end{equation}
\noindent
where $\delta n_{gd,Nk}$ and $\delta E_{gd,Nk}$ are the number 
and kinetic energy of particles dragged inwards from annulus $N$
to annulus $N-1$. 

We adopt the \citet{wet89} expression for velocity damping due to gas drag:

\begin{equation}
\frac{dV_{ik}}{dt} = \frac{-\pi C_D}{2m_{ik}} \rho_{g,i} V_{g,ik}^2 r_{ik}^2 ,
\end{equation}

\noindent
where $C_D$ = 0.5 is the drag coefficient and
$V_{gi} = (V_{ik} (V_{ik} + \eta_i))^{1/2}$ is 
the mean relative velocity of the gas.

\subsection{Poynting-Robertson Drag}

When the luminosity of the central star is large, the radiation
field can drag small particles towards the central star and can
circularize particle velocities relative to a circular orbit.
Radiation pressure can also eject very small particles from the disk.
\citet{bur79} have derived accurate expressions for these processes.
Because we do not treat the evolution of the smallest grains,
we ignore radiation pressure.  For Poynting-Robertson drag,
the inward drift of material is
\begin{equation}
\frac{\delta a_{ik}}{a_i} = \left ( \frac{2 + 3 e_{ik}^2}{(1 - e_{ik}^2)^{3/2}} \right ) ~ \frac{\eta_{pr} Q_{pr}}{a_i^2} ~ ,
\end{equation}
\noindent
where $Q_{pr}$ is the Mie scattering coefficient and
\begin{equation}
\eta_{pr} = \frac{2.53 \times 10^{11}}{\rho_{ik} r_{ik}} \left ( \frac{L_{\star}}{{\rm 1 ~ L_{\odot}}} \right ) ~ .
\end{equation}
\noindent
The mass density of an individual grain is $\rho_{ik}$.
The rate particles leave an annulus is
\begin{equation}
\frac{\delta n_{pr,ik}}{n_{ik}} = \frac{\delta a_{ik}}{a_i} ~ .
\end{equation}
\noindent
As with gas drag, we scale the number of particles lost by
annulus $N$ to calculate the number of particles added to 
this zone:
\begin{equation}
\delta n_{pr,Nk} = \left ( \frac{a_{N}}{a_{N+1}} \right )^2 \delta n_{pr,Nk}
\end{equation}
\noindent
For the kinetic energy:
\begin{equation}
\delta E_{pr,Nk}^{\prime} = \left ( \frac{a_{Nk}}{a_{N+1k}} \right )^3 ~ \delta E_{pr,Nk} ~ .
\end{equation}

\noindent
In these expressions, $\delta n_{pr,Nk}$ and $\delta E_{pr,Nk}$ are 
the number and kinetic energy of particles dragged inwards from 
annulus $N$ to annulus $N-1$. 

We also adopt the Burns et al. (1979) expression for horizontal velocity 
damping from Poynting-Robertson drag,
\begin{equation}
\frac{\delta h_{ik}^2}{\delta t^2} = -6.25 \left ( \frac{Q_{pr} h_{ik} \eta_{pr}}{a_i^2} \right )^2 ~ .
\end{equation}
\noindent
Poynting-Robertson drag does not change the vertical component of 
the velocity.

\subsection{Dust Luminosity and Optical Depth}

To calculate the dust luminosity, we follow \citet{ken99} and estimate 
the bolometric luminosity reprocessed by solid objects in the disk as 
$L/L_{\star}$ = $ \tau \Omega/4 \pi$, where $\tau$ is the radial 
optical depth and $\Omega$ is the solid angle of the disk or ring 
as seen from the central star.  The bolometric luminosity is the sum
of the thermal luminosity and the scattered light luminosity of the
dust grains.  For dust grains with albedo $\omega$, the 
scattered light luminosity is $\omega L$; the thermal luminosity 
is $(1 - \omega ) L$.  These approximations assume a grey opacity 
and are reasonable for $\tau \lesssim$ 1.  For a disk or ring geometry, 
the solid angle is $\Omega/4 \pi$ = $H/a$, where $H$ is the scale height 
and $a$ is the radial distance from the central star.

We calculate the radial optical depth through the grid to estimate the 
amount of absorption and scattering of starlight by dust particles in 
the grid.  We divide the optical depth $\tau$ through the grid into two 
pieces, $\tau_d$ measures the radial optical depth of bodies explicitly 
in the grid and $\tau_s$ measures the radial optical depth of smaller bodies.  
We adopt the geometric optics limit for bodies with $r_k \gg \lambda$ 
where $\lambda$ is the wavelength of observation.  For the large bodies 
in the grid:
\begin{equation}
\tau_d = \sum_{i=1}^N \sum_{k=2}^M N_{ik} ~ \sigma_{ik} ~ \Delta a_i
\end{equation}
where $N_{ik}$ is the number density of mass batch $k$ in annulus 
$i$ and $\sigma_{ik}$ is the extinction cross-section.  We adopt
$\sigma_{ik} = 2 \pi r_{ik}^2$ and a volume 
$V_{ik} = 4 \pi a_i \Delta a_i H_{ik}$.
The optical depth is then independent of the width of the annulus:
\begin{equation} 
\tau_d = \sum_{i=1}^N \sum_{k=2}^M \frac{n_{ik} ~ r_{ik}^2}{2 ~ a_i ~ H_{ik}} ~ . 
\end{equation}

For the smaller bodies, we must make assumptions about the variation
of $n$ and $H$ with particle size.  In most calculations, the vertical
scale height of the dust $H_{ik}$ is either constant or grows slowly
for smaller particles.  Adopting a constant $H_{ik}$ thus maximizes the 
optical depth of the smallest grains.  The cumulative number density usually
follows $N_C \propto r^{-\beta}$ with $\beta \approx$ 2.5 for calculations
with fragmentation \citep{doh69,wil94,tan96}.  
Gas and Poynting-Robertson drag preferentially
remove the smallest particles from the grid;  these processes should
lower the exponent in the number density to $\beta \approx$ 1.5 when
either $\rho_{g}$ or $L_\star$ is large.  Several test calculations 
confirm this dependence.  We thus consider two cases,
$\beta = 2.5$ -- when small particles dominate the opacity and
$\beta = 1.5$ -- when large particles dominate the opacity.
The optical depth is then
\begin{equation}
\tau_s = \sum_{i=1}^N \sum_{k=1}^{-\infty} \frac{n_{i1} ~ r_{i1}^2}{2 ~ a_i ~ H_{i1}} r_{ik}^{1/2} ~ = K_{\tau} \sum_{i=1}^N \frac{n_{i1} ~ r_{i1}^2}{2 ~ a_i ~ H_{i0}}
\end{equation}
\noindent
where
\begin{equation}
K_\tau = \left\{ \begin{array}{l l l}
         3 & \hspace{5mm} & \beta = 1.5 \\
         500 ~ (r_{min} / 1 ~ \mu m)^{-1/2} & \hspace{5mm} & \beta = 2.5 \\
         \end{array}
	 \right .
\end{equation}
\noindent
Radiation pressure limits the size of the smallest particle to 
radii $\sim$ 1--10 $\mu$m.  We plan to address the large range
in possible optical depths in future studies where we explicitly
calculate the evolution of small particles.

With these definitions for the optical depth, the bolometric luminosity
from planetesimals in the grid is
\begin{equation} 
L_d = \sum_{i=1}^N \sum_{k=2}^M \frac{n_{ik} ~ r_{ik}^2}{2 ~ a_i^2} ~ . 
\end{equation}
\noindent
This expression is independent of the scale height $H$.
The bolometric luminosity from small particles in the grid is
\begin{equation}
L_{min} = 3 \sum_{i=1}^N \frac{n_{i1} ~ r_{i1}^2}{2 ~ a_i^2}
\end{equation}
\noindent
in the low optical depth limit and
\begin{equation} 
L_{max} = 500 ~  (r_{min} / 1 ~ \mu m)^{-1/2} \sum_{i=1}^N \frac{n_{i1} ~ r_{i1}^2}{2 ~ a_i^2}
\end{equation}
\noindent 
in the high optical depth limit.

\citet{kri00} 
model the dynamical
evolution of two small grain populations in debris disks. They derive 
size distributions for grains with $r_{ik} \lesssim$ 1 mm for various 
assumptions for the collisional evolution of larger grains. 
Their calculations do not include velocity evolution.  \citet{kri00}
show that collisions dominate Poynting-Robertson drag in debris disks 
with substantial optical depth, $\tau \gtrsim 10^{-4}$.  If supported
by calculations with velocity evolution, these results suggest that
our maximum optical depth is appropriate for the early evolution of
debris disk systems.

\subsection{Tests of the Evolution Code}

\citet{kb01} describe various successful attempts to match results derived 
from the multi-annulus code with published calculations of velocity evolution.
We have fully tested the new code in a variety of situations to verify
that our algorithms conserve mass, angular momentum, and kinetic energy
throughout an evolutionary sequence.  The code reproduces previous results
on particle growth in the Kuiper Belt \citep{kl99}.  Comparisons with 
previously published calculations of terrestrial planet formation 
\citep[e.g.,][]{wei97} will be described in a forthcoming publication.

\clearpage


\epsfxsize=6.0in
\epsffile{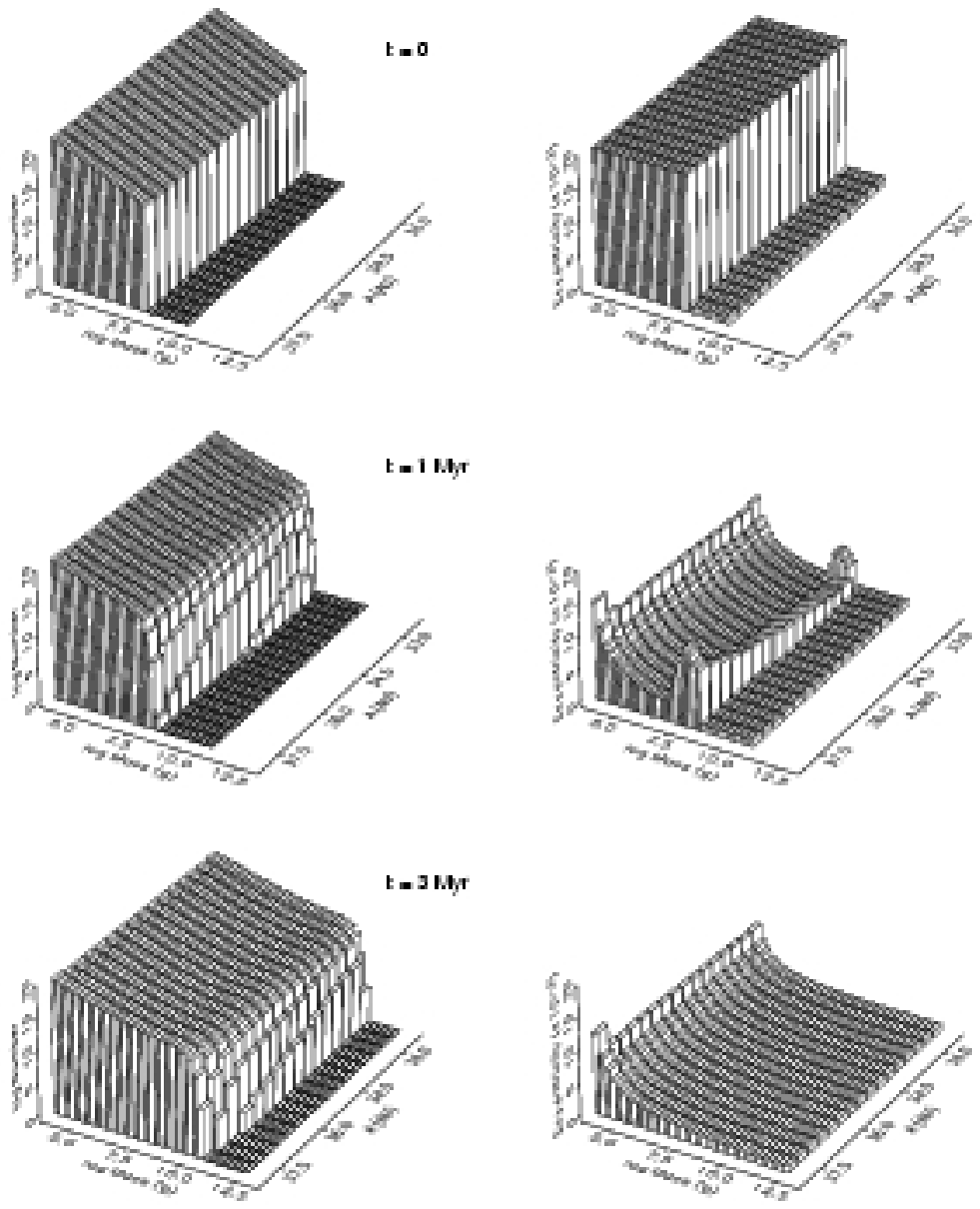}
\figcaption[Kenyon.fig1.ps]
{Evolution of particle number (left panels) and eccentricity 
(right panels) for planetesimals in orbit around a 3 \msun~star 
at 35 AU.  The sixteen annuli in the grid for this ring model 
contain planetesimals with $r_0$ = 0.1--10 m, $e_0 = 
2 \times 10^{-2}$, $N(m) \propto m^{-1}$, and a total mass
of 0.667 $M_{\oplus}$ at $t = 0$ (top two panels).  The
middle panels show the particle number and eccentricity at
$t$ = 1 Myr; the bottom panels show the particle number and 
eccentricity at $t$ = 3 Myr.  Despite considerable mass loss 
from shattering, collisional damping reduces the orbital 
eccentricities to $\sim$ $10^{-3}$ on timescales of 1--10 Myr.  
Low velocity collisions then promote growth of larger objects 
through mergers. Style adapted from Weidenschilling et al. (1997).}

\epsfxsize=9.0in
\hskip -10ex
\epsffile{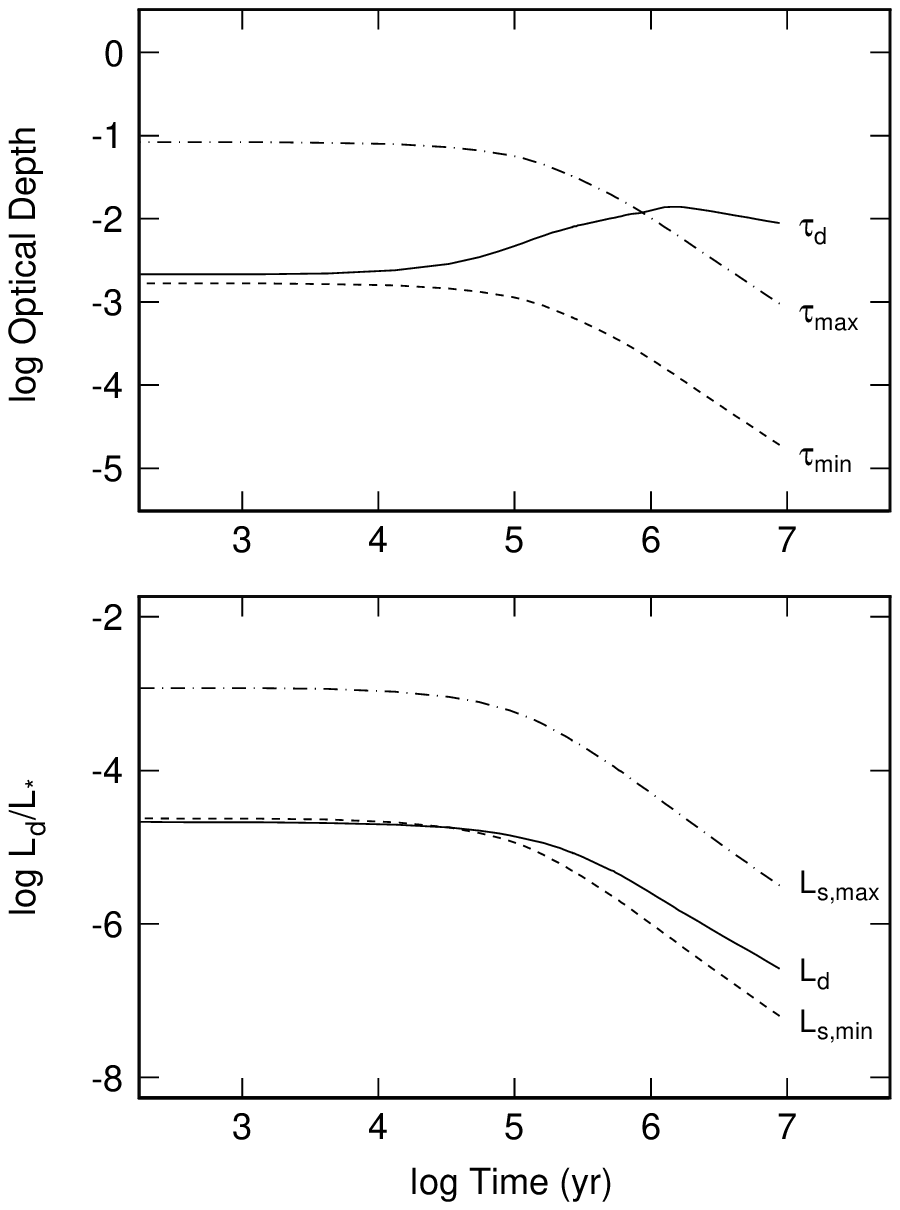}
\figcaption[Kenyon.fig2.eps]
{Evolution of optical depth (upper panel) and reprocessed luminosity
(lower panel) for the ring model of Figure 1.}

\epsfxsize=9.0in
\hskip -10ex
\epsffile{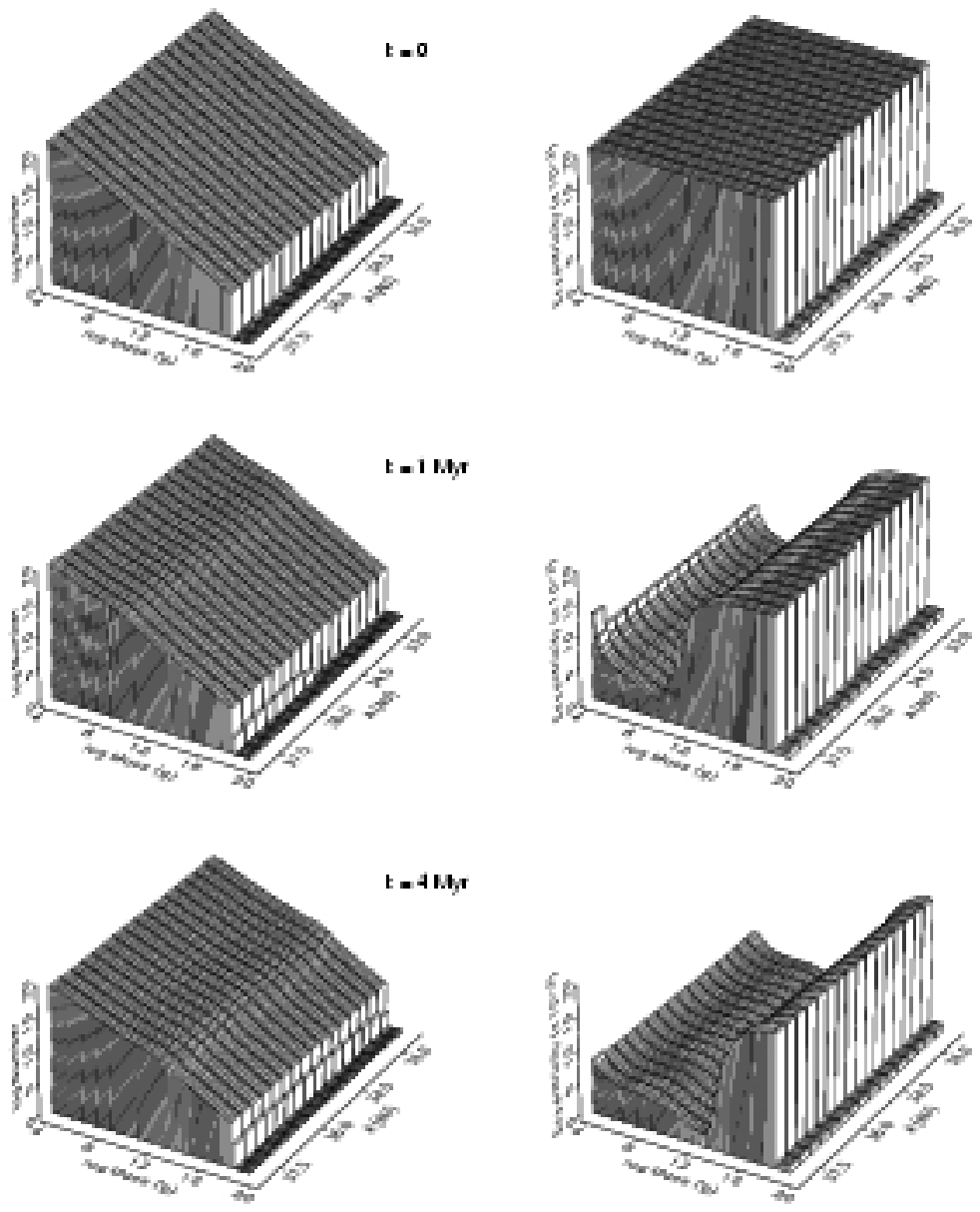}
\figcaption[Kenyon.fig3.eps]
{As in Figure 1, for planetesimals with $r_{max}$ = 10 km
at $t = 0$ (top panels),
$t$ = 1 Myr (middle panels), and
$t$ = 4 Myr (bottom panels.}

\epsfxsize=8.0in
\hskip -10ex
\epsffile{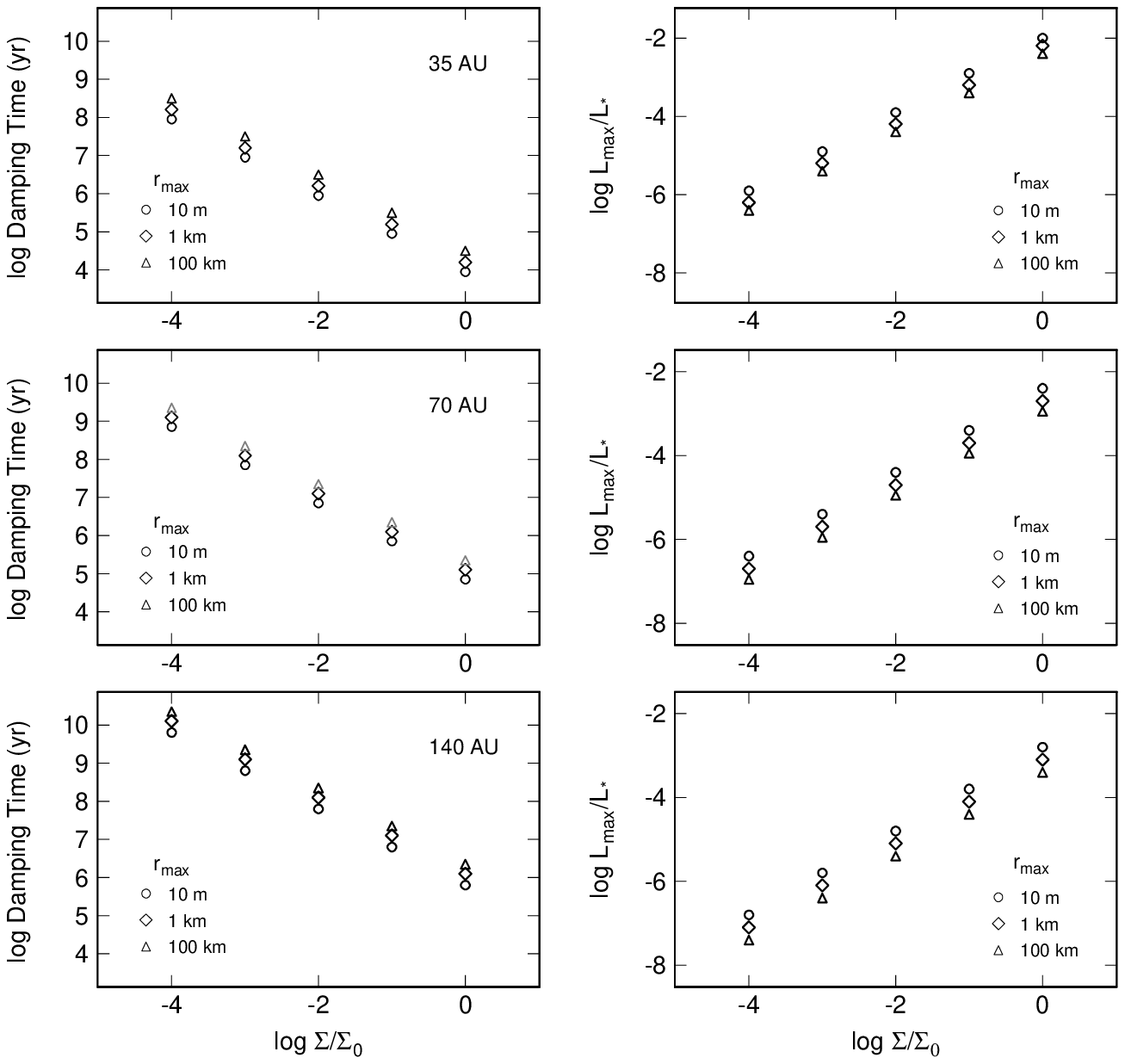}
\figcaption[Kenyon.fig4.eps]
{Damping time (left panels) and dust luminosity (right panels) as a
function of initial surface density for ring models at 35, 70, 
and 140 AU. Models with $\Sigma = \Sigma_0$ have surface densities of
solid material similar to a minimum mass solar nebula. All models have 
$r_{min}$ = 10 cm; model results for different values of $r_{max}$ are 
shown as indicated by the legend in each panel.}

\epsfxsize=8.0in
\hskip -10ex
\epsffile{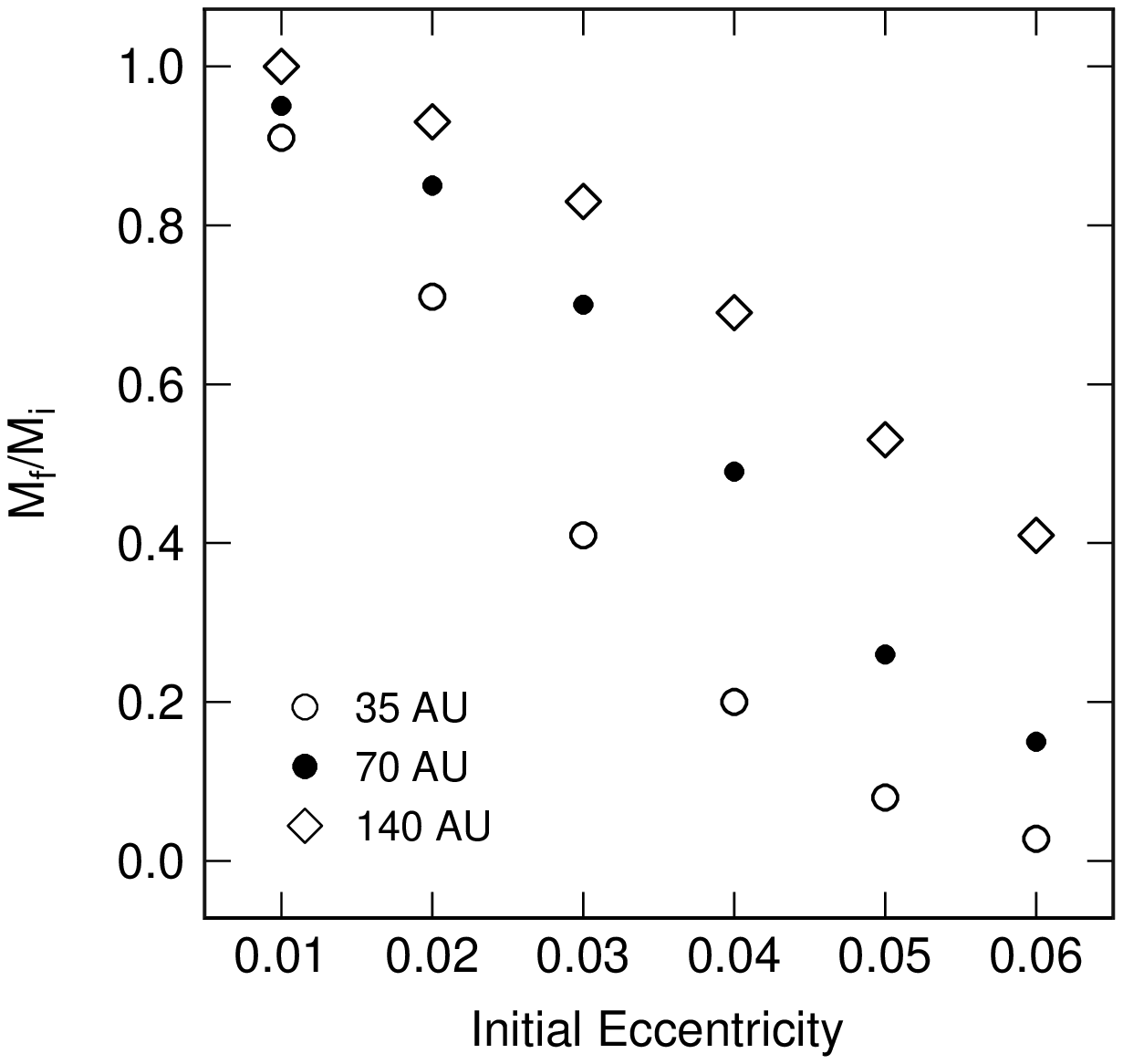}
\figcaption[Kenyon.fig5.eps]
{The ratio of final mass, $M_f$, to the initial mass, $M_i$, as a 
function of the initial eccentricity for ring models at 35 AU
(open circles), 70 AU (filled circles), and 140 AU (open diamonds).
Models with large $e_0$ lose most of their mass.}

\epsfxsize=9.0in
\hskip -10ex
\epsffile{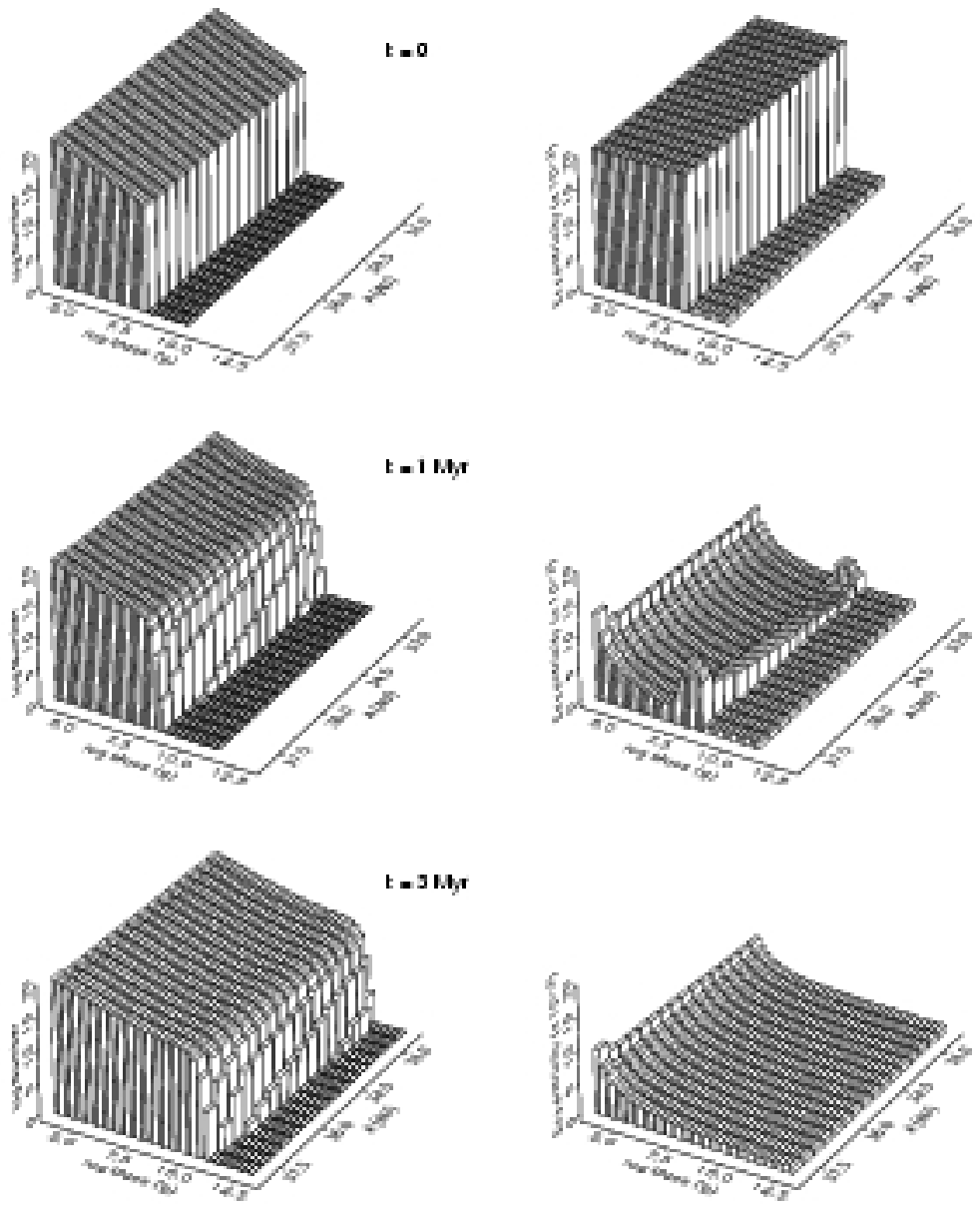}
\figcaption[Kenyon.fig6.eps]
{As in Figure 1, for a ring model with gas drag. The initial gas to dust 
mass ratio is 10:1.}

\epsfxsize=9.0in
\hskip -10ex
\epsffile{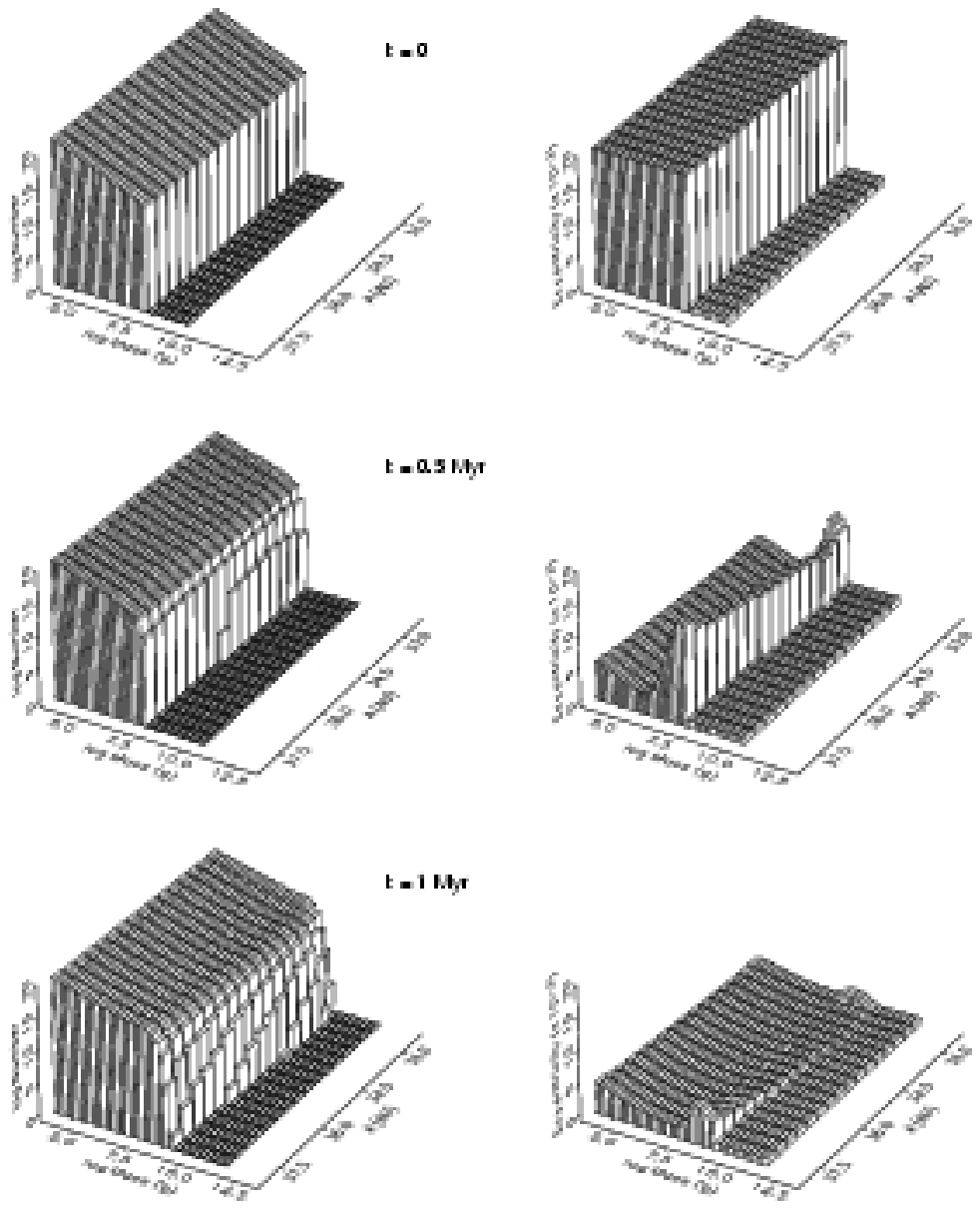}
\figcaption[Kenyon.fig7.eps]
{As in Figure 6, for a ring model with an initial gas to dust mass ratio of 100:1.}

\epsfxsize=8.5in
\epsffile{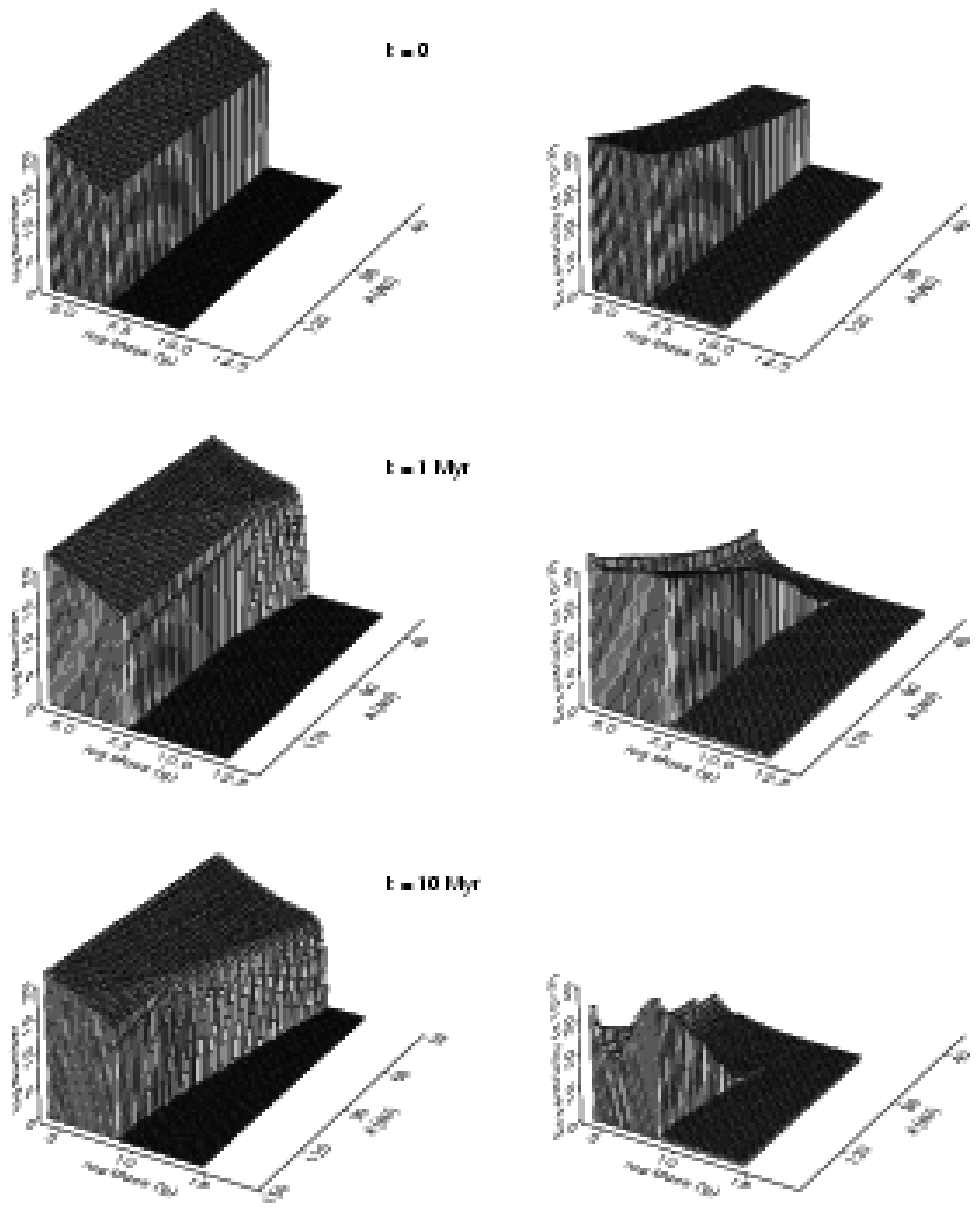}
\figcaption[Kenyon.fig8.ps]
{Evolution of particle number (left panels) and eccentricity 
(right panels) for planetesimals in orbit around a 3 \msun~star 
at 30--150 AU.  The 64 annuli in the model grid contain
planetesimals with $r_0$ = 0.1--10 m, $e_0 = 
2 \times 10^{-2}$, $N(m) \propto m^{-1}$, and a total mass
of 10 $M_{\oplus}$ at $t = 0$ (top two panels). The model grid 
extends from 30 AU at the inner edge to $\sim$ 150 AU at the outer 
edge.  Despite considerable mass loss from shattering, collisional
damping reduces the orbital eccentricities to $\sim$ $10^{-3}$
on timescales of 1--10 Myr.  Low velocity collisions then
promote growth of larger objects through mergers. }

\epsfxsize=8.5in
\hskip 10ex
\epsffile{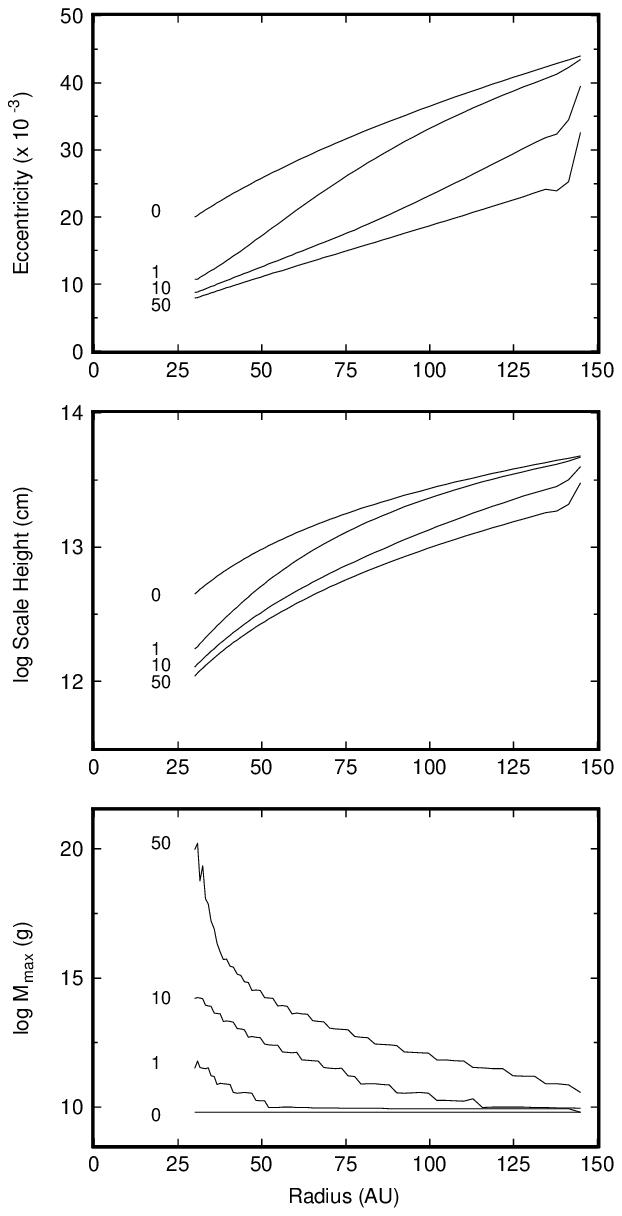}
\figcaption[Kenyon.fig9.ps]
{Evolution of the smallest and largest bodies in the full disk model 
of Figure 8.
Top panel: eccentricity of the smallest bodies, $r_k$ = 10 cm, as a
function of disk radius at $t$ = 0, 1, 10, and 50 Myr.  The abrupt
rise in $e$ at the outer edge of the grid is a numerical artifact.
Middle panel: scale height of bodies with $r_k$ = 10 cm as a
function of disk radius at $t$ = 0, 1, 10, and 50 Myr.
Bottom panel: mass of the largest body in each annulus at $t$ = 0, 1, 10,
and 50 Myr.  Large bodies at the inner edge of the grid are starting the
runaway growth phase just as growth begins at the outer edge of the grid.}

\epsfxsize=8.0in
\hskip -10ex
\epsffile{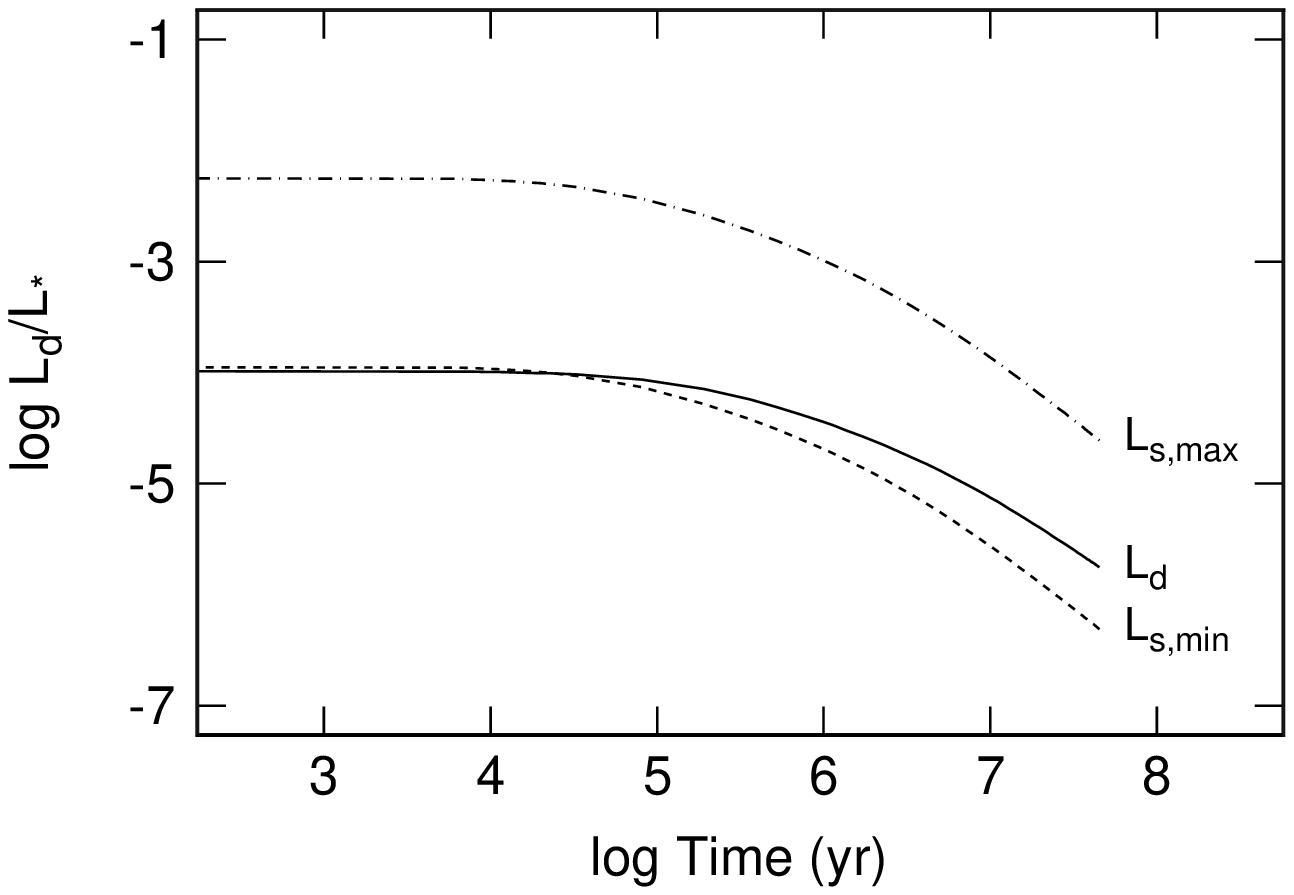}
\figcaption[Kenyon.fig10.ps]
{Evolution of dust luminosity for the full disk model of Figure 8.}

\epsfxsize=7.0in
\hskip -10ex
\epsffile{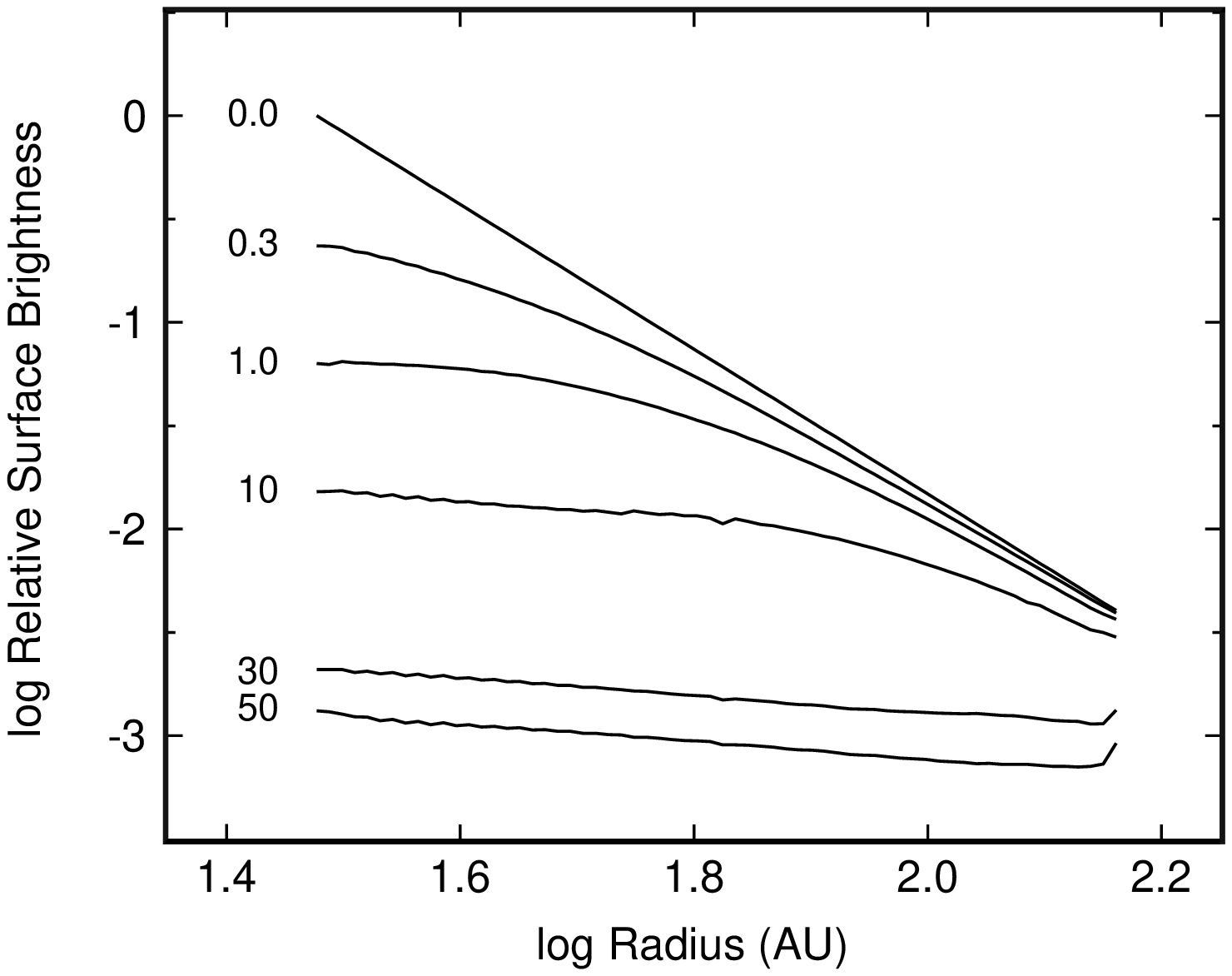}
\figcaption[Kenyon.fig11.ps]
{Evolution of surface brightness for the full disk model of Figure 8.
At $t = 0$, the surface brightness profile is a power law,
$I \propto a_i^{-p}$, with $p = 3.5$.  For $t$ = 0--1 Myr,
the surface brightness at the inner edge of the disk fades 
by an order of magnitude; the surface brightness at the outer 
edge, log $r_i \gtrsim$ 1.8--2.0, barely changes.  The outer 
edge of the disk begins to fade at $t$ = 1--10 Myr.  For $t$ 
= 10--50 Myr, the entire disk fades by an order of magnitude.
and begins to produce planets. For $t$ = 30 Myr and 50 Myr,
the rise in surface brightness at the outer edge of the disk 
is a numerical artifact.}

\clearpage

\hskip -10ex
\epsffile{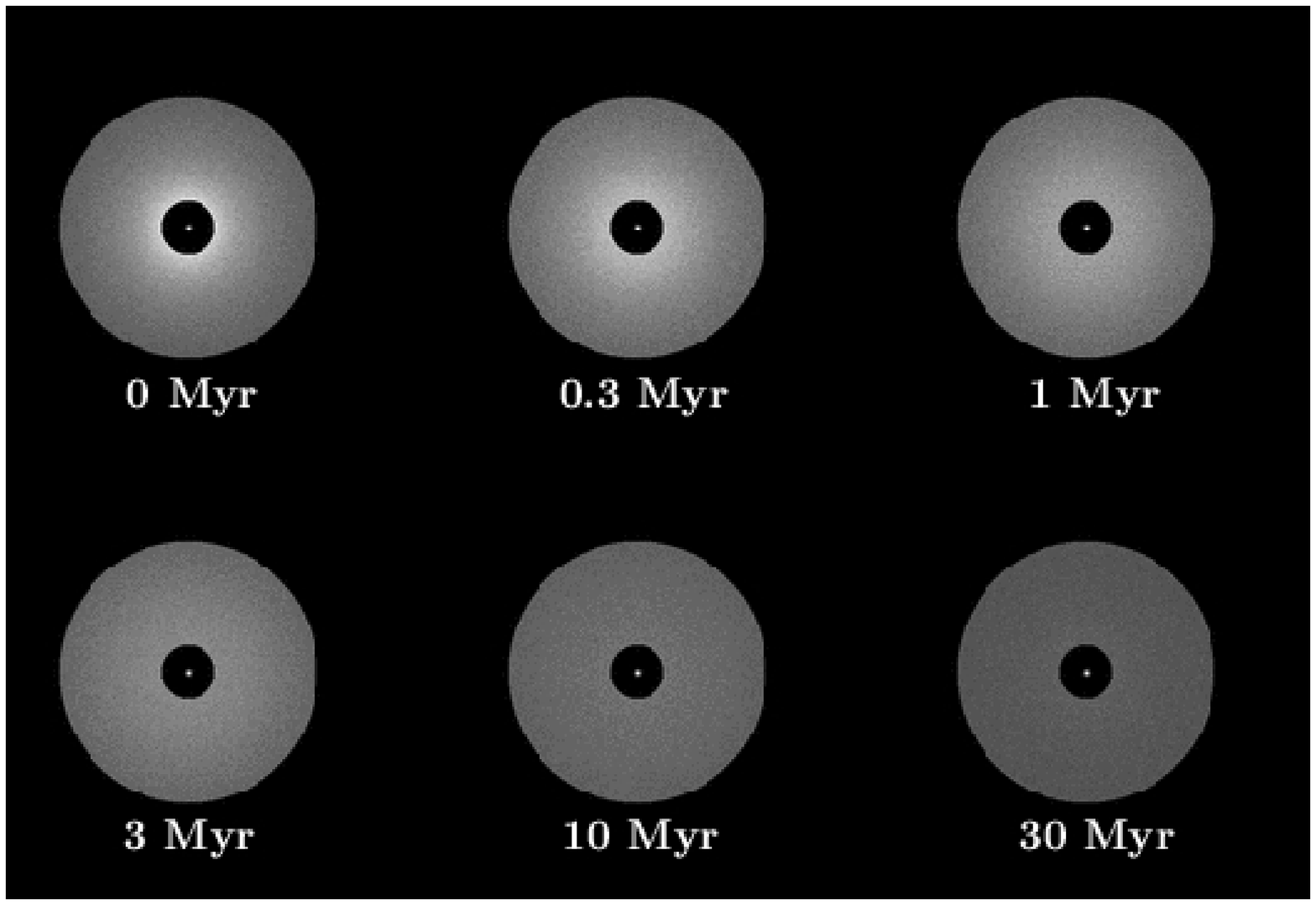}
\figcaption[Kenyon.fig12.ps]
{Model images for the full disk calculation of Figure 8.  The panels 
show surface brightness distributions in the disk at selected times 
in the calculation.}

\end{document}